\documentclass[preprint]{aastex}

\usepackage{natbib}
\usepackage{color}
\newcommand{\update}[2]{#1}

\newcommand{\Msun}{M$_\odot$}

\newcommand{\mum}{$\mu$m}

\newcommand{\etal}{et~al.}

\begin{document}
\title{Spitzer Observations of IC 2118}

\slugcomment{Version from \today}

\author{S.\ Guieu\altaffilmark{1,2},
  L.\ M.\ Rebull\altaffilmark{1}, 
  J.\ R.\ Stauffer\altaffilmark{1},
  F.\ J.\ Vrba\altaffilmark{3},
  A.\ Noriega-Crespo\altaffilmark{1},
T.\ Spuck\altaffilmark{4}, 
T.\ Roelofsen Moody\altaffilmark{5}, 
B.\ Sepulveda\altaffilmark{6}, 
C.\ Weehler\altaffilmark{7},
A.\ Maranto\altaffilmark{8},  
D.\ M.\ Cole\altaffilmark{1}, 
N.\ Flagey\altaffilmark{1},
R.\ Laher\altaffilmark{1}, 
B.\ Penprase\altaffilmark{9}, 
S.\ Ramirez\altaffilmark{10},  
S.\ Stolovy\altaffilmark{1} 
}

\altaffiltext{1}{Spitzer Science Center/Caltech, M/S 220-6, 1200
E.\ California Blvd., Pasadena, CA  91125}
\altaffiltext{2}{Current addres:  European Southern Observatory, Alonso de
C\'ordova 3107, Casilla 19001, Vitacura, Santiago 19, Chile e-mail:
sguieu@eso.org}
\altaffiltext{3}{U.S. Naval 
Observatory, Flagstaff Station, 10391 West Naval Observatory Rd., 
Flagstaff, AZ 86001-8521}
\altaffiltext{4}{Oil City Area Senior High School, Oil City, PA}
\altaffiltext{5}{Bassick High School, Bridgeport, CT; currently New
Jersey Astronomy Center, Raritan Valley Community College, Somerville,
NJ}
\altaffiltext{6}{Lincoln High School, Stockton, CA}
\altaffiltext{7}{Luther Burbank High School, San Antonio, TX}
\altaffiltext{8}{McDonogh School, Owings Mills, MD}
\altaffiltext{9}{Pomona College, CA}
\altaffiltext{10}{NASA Exoplanet Science Institute, IPAC, M/S 1002-22, 
1200 E.\ California Blvd., Pasadena, CA 91125}

\begin{abstract}

IC~2118, also known as the Witch Head Nebula, is a wispy, roughly
cometary, $\sim$5 degree long reflection nebula, and is thought to be
a site of triggered star formation. In order to search for new young
stellar objects (YSOs), we have observed this region in 7 mid- and
far-infrared bands using the Spitzer Space Telescope and in 4 bands in
the optical using the U.~S.~Naval Observatory 40-inch telescope.  We
find infrared excesses in 4 of the 6 previously-known T~Tauri stars in
our combined infrared maps, and we find 6 entirely new candidate YSOs,
one of which may be an edge-on disk.  Most of the YSOs seen in the
infrared are Class II objects, and they are all in the ``head'' of the
nebula, within the most massive molecular cloud of the region.

\end{abstract} 
   \keywords{stars: formation -- stars: circumstellar matter --
stars: pre-main sequence -- ISM: clouds -- ISM: individual (IC 2118)
-- infrared: stars -- infrared: ISM}

\section{Introduction}
\label{sec:intro}

Our general understanding of how and why it is stars form rests upon
the answers to a set of more basic questions: given a particular
locale, how, why, and when did stars form there? Current theory tells
us that in some regions, stars simply collapse under the influence of
their own natal clouds' gravity, but in others, the process may be
heavily influenced by external events such as nearby supernovae, which
may prompt parts of the cloud to form stars. Many people have searched
for evidence of triggered star formation on large and small scales
(e.g., \citealt{Briceno-2007} and references therein).  In the cases
where triggered star formation is suspected, a complete inventory of
the young stars formed by the cloud is useful specifically to (a)
search for evidence of sequential star formation within the cloud and
with respect to other nearby clouds, and (b) within the individual
cloud, understand the initial mass function (IMF), the total mass in
the cloud vs.\ stars, and the star formation efficiency.  The former
strengthens (or weakens) the case for having found legitimate
triggered star formation, and the latter allows closer study of the
initial conditions of star formation (e.g.,
\citealt{Ballesteros-2007}).  If the specific cloud does turn out to
be forming stars as a result of triggering, then a detailed comparison
of the IMF and star formation efficiency for a triggered cloud versus
a cloud that has collapsed under its own gravitational influence  will
shed further light on the mechanics of star formation itself.

Finding all of the young stars in any given cloud in the optical can
be problematic due to extinction from the cloud itself.  However, the
Spitzer Space Telescope \citep{Werner-2004} provides an excellent
platform for surveying star-forming regions in the mid-IR and far-IR,
enabling stars with IR excesses to be identified in a relatively
straightforward fashion, even those embedded fairly deeply in the
cloud.  Because Spitzer efficiently maps large regions of sky, it is a
particularly useful tool for surveying clouds subtending relatively
large areas.

\begin{figure*}
\includegraphics[width=\hsize]{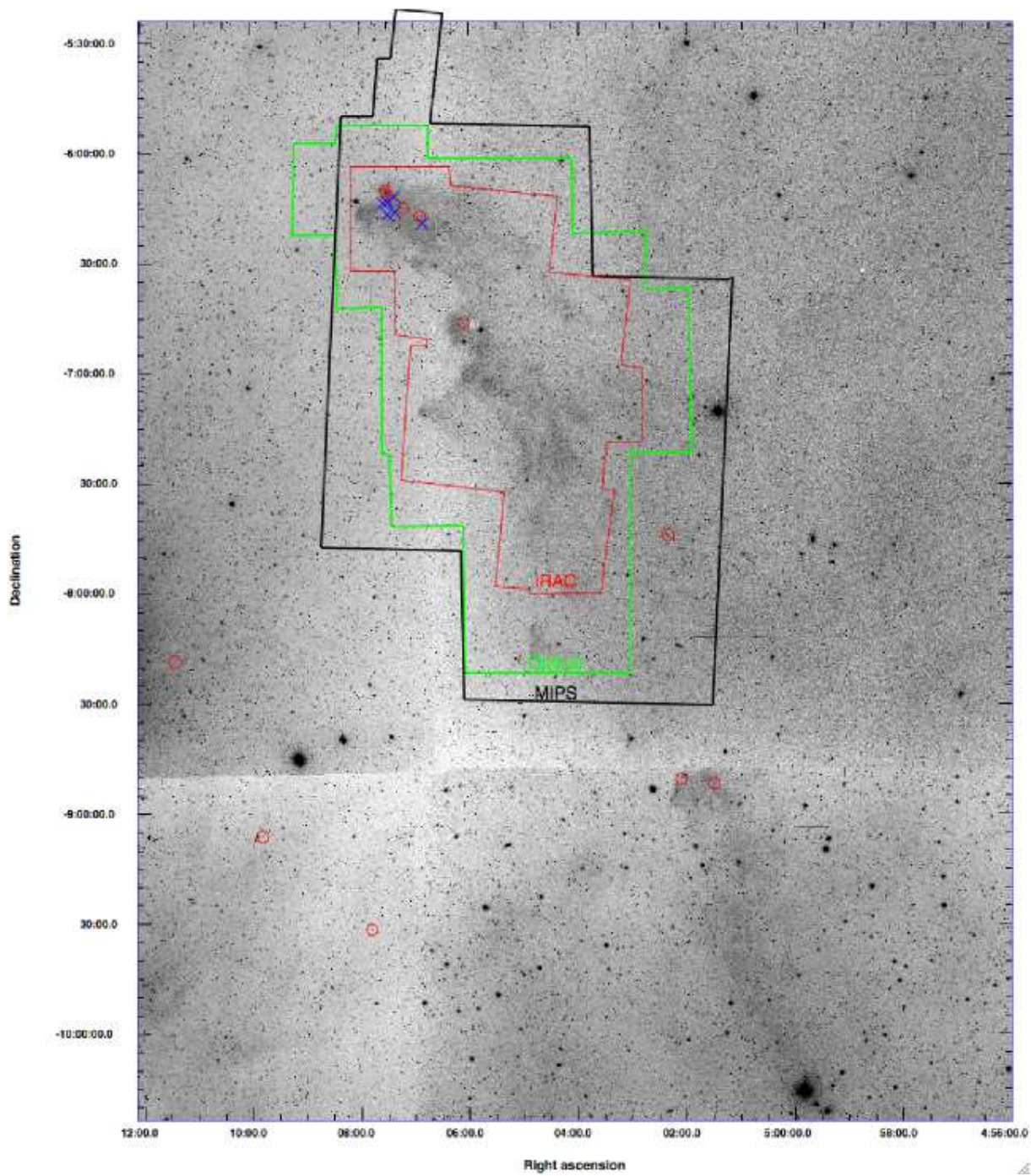}
\caption{Palomar Observatory Sky Survey (POSS) red image of IC~2118,
with the eleven previously-identified T~Tauri stars indicated (red circles),
as well as the six new candidate YSOs found here (blue $\times$).
The outlines of our IRAC (red), MIPS (black), and optical (green)
observations are included
for reference.
\label{fig:where}}
\end{figure*}

IC~2118 \citep{Kun-2001, Kun-2004}, the Witch Head Nebula, is $\sim$5
degrees long; its ``wind-blown'' appearance is similar in the optical
and infrared -- see Figure~\ref{fig:where} for the wide-field optical
image from the Palomar Observatory Sky Survey. It has been cited as an
example of triggered star formation. The specific distance to the
object is somewhat controversial.  If it is close to $\sim$400 pc, the
trigger could be the Trapezium \citep{Lee-2007}.  If, instead, it is
closer, some $\sim$200 pc away, the possible trigger is the
Orion-Eridanus superbubble \citep{Kun-2001, Kun-2004}, and the cloud
is likely to be additionally excited by $\beta$ Orionis (Rigel, at 264
pc; \citealt{van-Leeuwen-2007}).  It is possible that interactions
with both the expanding superbubble and the winds from Rigel (both to
the East of IC 2118) create the cometary shape of the nebula.  For the
rest of this paper, we take the distance to IC 2118 to be 210 pc
\citep{Kun-2001}, although we comment more on this distance
uncertainty below.

Some previous authors have noted star formation in this region;
\cite{Kun-2001} summarizes all previously identified objects here,
including red nebulous objects found by \cite{Cohen-1980}.   With a
$^{12}$CO survey of the IC~2118 region, \cite{Kun-2001} detected six
molecular clouds which include MBM~21 and MBM~22 (G~208.4-28.3 and
G~208.1-27.5, respectively). Kun and collaborators \citep{Kun-2001,
Kun-2003, Kun-2004}  conducted an objective prism survey with
spectroscopic follow-up to search for new T~Tauri stars in the IC~2118
region.  They concluded that there are 11 young stars in the vicinity,
8 of which they discovered, and 6 of which are located within
(projected onto) the higher-density nebulosity of the cloud; see
Table~\ref{tab:prevknown} (where all the types come from
\citet{Kun-2004}) and Figure~\ref{fig:where}.  Based on this census of
YSOs in IC~2118, the star formation efficiency here was taken to be
$\sim$3\%.

We used Spitzer to survey the heart of the IC~2118 region, with the
Infrared Array Camera (IRAC; \citealt{Fazio-2004}) at 3.6, 4.5, 5.8,
and 8 microns, and the Multiband Imaging Photometer for Spitzer (MIPS;
\citealt{Rieke-2004}) at 24, 70, and 160 microns.  In this paper, we
use these data to search for candidate young stellar objects (YSOs)
with infrared excesses.  We use our additional optical and Two-Micron
All-Sky Survey (2MASS; \citealt{Skrutskie-2006}) photometric data in
an effort to segregate YSOs from background galaxies.  While one
previously-known T~Tauri star appears without an excess in our MIPS
map alone, we find infrared excesses in 4 of the 5 previously-known
T~Tauri stars in our combined IRAC \& MIPS maps, and we find 6
entirely new candidate YSOs.  

The observations and data reduction are described in 
\S\ref{sec:obs}.  We select YSO candidates using Spitzer colors in
\S\ref{sec:findthem}, and discuss their overall properties in
\S\ref{sec:properties}.  Finally, we discuss some wider implications
and summarize our main points in \S\ref{sec:concl}.  

\begin{deluxetable}{lccll}
\tablecaption{Previously-identified T~Tauri stars in
IC~2118\label{tab:prevknown}}
\tabletypesize{\tiny}
\tablewidth{0pt}
\tablehead{
\colhead{name}  &  \colhead{RA (J2000)} & \colhead{Dec (J2000)} &
\colhead{spectral type\tablenotemark{a}} & \colhead{other names} }
\startdata
IRAS 04591-0856        &   05 01 30.2 & -08 52 14& \ldots& HHL 17, G13\\ 
2MASS 05020630-0850467 &   05 02 06.3 & -08 50 47& M2 IV & \ldots \\ 
RXJ 0502.4-0744\tablenotemark{b}        &   05 02 20.8 & -07 44 10& \ldots& 2MASS 05022084-0744099\\ 
2MASS 05060574-0646151\tablenotemark{c} &   05 06 05.7 & -06 46 15& G8:   &  (may not be a member of IC~2118; see \citealt{Kun-2004})\\ 
2MASS 05065349-0617123\tablenotemark{c,d} &   05 06 53.5 & -06 17 12& K7 IV & \ldots\\ 
2MASS 05071157-0615098\tablenotemark{c,d} &   05 07 11.6 & -06 15 10& M2 IV & IRAS F05047-0618\\ 
2MASS 05073016-0610158\tablenotemark{c,d} &   05 07 30.2 & -06 10 16& K6 IV\tablenotemark{e} & IRAS 05050-0614, Kiso H$\alpha$ A0974-19, RNO 37\\ 
2MASS 05073060-0610597\tablenotemark{c,d} &   05 07 30.6 & -06 11 00& K7 IV\tablenotemark{e} & Kiso H$\alpha$ A0974-20, RNO 37\\ 
RXJ 0507.8-0931        &   05 07 48.3 & -09 31 43& \ldots& 2MASS 05074833-0931432\\ 
2MASS 05094864-0906065 &   05 09 48.6 & -09 06 07& G8    & (may not be a member of IC~2118; see \citealt{Kun-2004})\\ 
2MASS 05112460-0818320 &   05 11 24.6 & -08 18 32& M0    & \ldots \\
\enddata
\tablenotetext{a}{From Kun et al.~(2004).}
\tablenotetext{b}{Covered only by our MIPS map.}
\tablenotetext{c}{Covered by our IRAC and MIPS maps.}
\tablenotetext{d}{Has a mid-IR excess as determined here via Spitzer colors.}
\tablenotetext{e}{\citet{Lee-2005} identifies 2MASS 05073016-0610158
as an M0, and 2MASS 05073060-0610597 as an M2.}
\end{deluxetable}

\section{Observations and data reduction}
\label{sec:obs}

\subsection{Spitzer Observations}

\begin{figure*}
\includegraphics[width=\hsize]{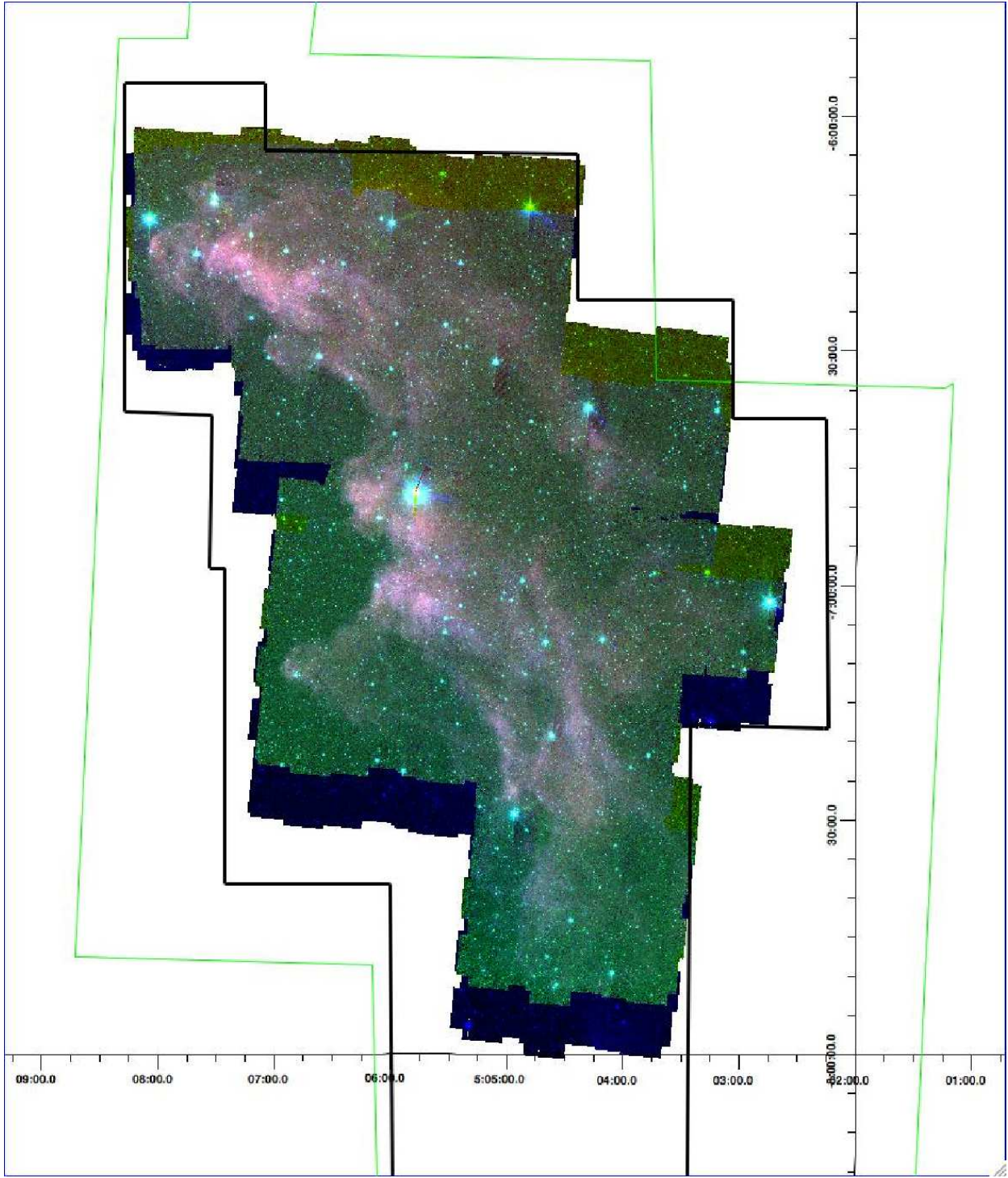}
\caption{Three-color view of IC~2118 with 3.6~$\mu$m in blue,
4.5~$\mu$m in green and 8~$\mu$m in red. Outlines of our optical
(black line) and MIPS (green line) surveys are also plotted.
\label{fig:mosaics}} 
\end{figure*}

We observed IC 2118 with Spitzer using both IRAC and MIPS. These
observations were obtained as director's discretionary time (DDT) as
part of the Spitzer Observing Program for Teachers and Students.  Most
of the data were part of programs 235 and 266 (PI: T.~Spuck), obtained
in Spring 2005 and 2006, respectively; additional observations of
small portions of the map were obtained as part of program 462 (PI:
L.~Rebull) in Fall 2007 and Spring 2008 in an effort to increase the
legacy value of the data set.  The IRAC observation was broken into
nine astronomical observation requests (AORs), and the MIPS
observation into three; the associated AORKEYs\footnote{AORKEYs are
the unique 8-digit identifier for the AOR, which can be used to
retrieve these data from the Spitzer Archive.}  are given in Table
\ref{table:aors}. The IRAC observations used the 12-second
high-dynamic-range (HDR) mode such that each exposure consists of a
0.6 sec and a 12 sec frame.  There were three dithers per position to
minimize instrumental effects, stepping by 260$^{\prime\prime}$ for
each portion of the map. The IRAC observation covers $\sim$1.6~deg$^2$
in all four IRAC bands (3.6, 4.5, 5.8, and 8 $\mu$m);
Figure~\ref{fig:mosaics} shows a three-color IRAC mosaic made using
4.5, 5.8 and 8 $\mu$m.  The outlines of our MIPS and optical
observations are included for reference.  

\begin{figure*}
\includegraphics[height=0.75\hsize,angle=90]{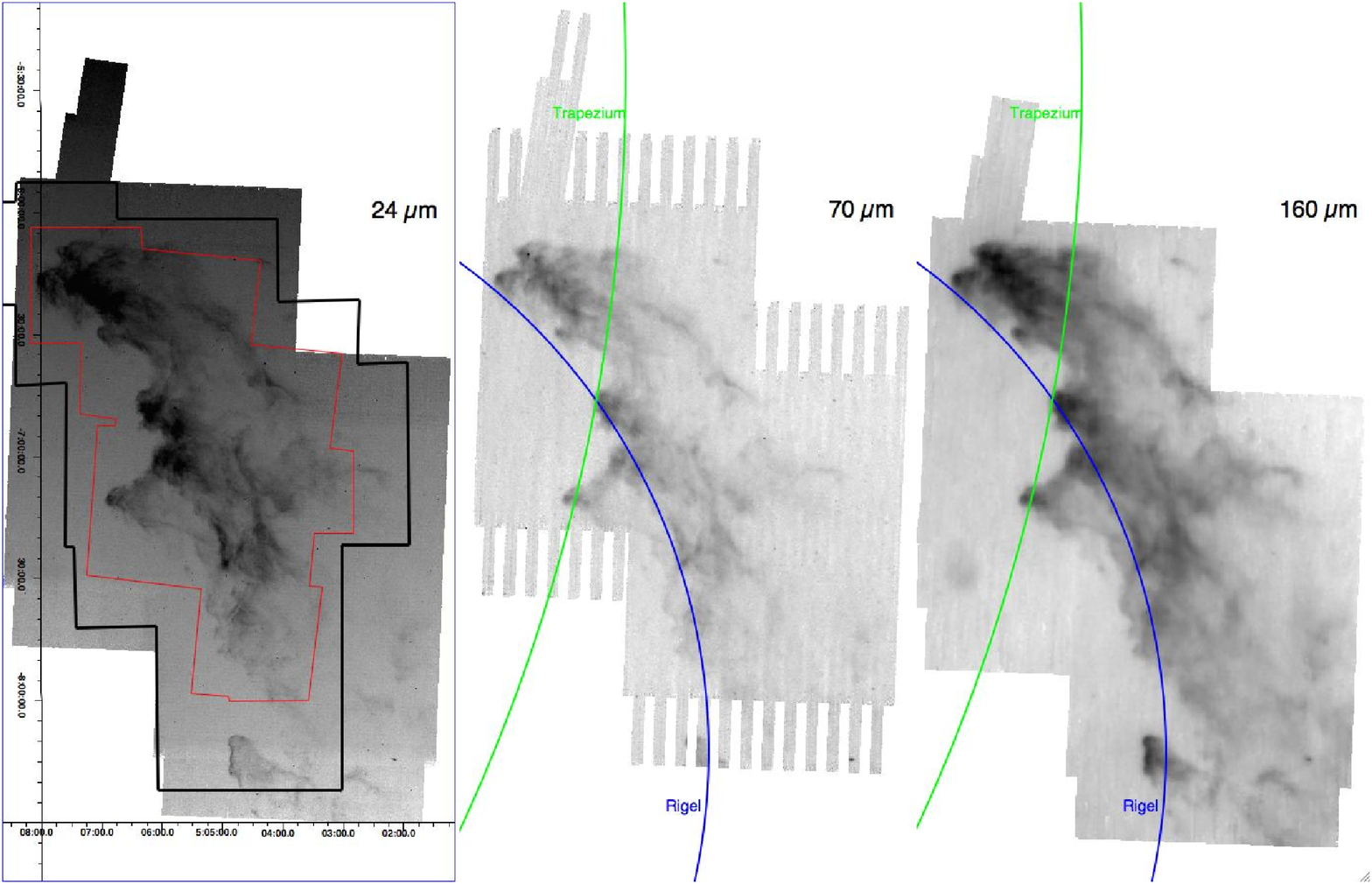}
\caption{The 24~$\mu$m, 70~$\mu$m and 160~$\mu$m MIPS mosaics of
IC~2118.  Outlines of our optical (black line) and IRAC (red line)
surveys are also included on the 24~$\mu$m mosaic. 
The arcs are from circles that are centered on the Trapezium 
(green circle, 7.35$\arcdeg$ radius) and on Rigel (blue circle, 
4.48$\arcdeg$ radius). 
\label{fig:mosaic24}}
\end{figure*}

The MIPS observations were fast scans, and each scan leg is offset by
160$^{\prime\prime}$ (half an array). There is complete coverage at 70
$\mu$m and two scan legs per position at 24 $\mu$m. We did not
anticipate having viable 160 \mum\ data, so we did not plan on
complete coverage at this bandpass.  However, the observations were
conducted in cold MIPS campaigns, which results in viable data even if
the coverage is incomplete. The MIPS observation covers $\sim$4
deg$^2$; the 24, 70, and 160 $\mu$m full mosaics are shown in
Figure~\ref{fig:mosaic24}.

\subsection{IRAC Data Reduction}

For the IRAC data, we started with the Spitzer Science Center (SSC)
pipeline-processed basic calibrated data (BCDs), version S18.7 for 
program 235 and S18.5 for the two other programs; for our purposes,
there is no difference between these pipeline versions. We further
processed the BCDs with the IRAC artifact mitigation software written
by S.~Carey and available on the SSC web site. Mosaics were
constructed from the corrected BCDs using the SSC mosaicking and point
source extraction software package MOPEX \citep{Makovoz-2005}. The
mosaics were computed to have a pixel scale of 1$^{\prime\prime}$.22
pixel$^{-1}$.  

We ran the APEX portion of MOPEX for source detection
on the mosaics;  we performed aperture photometry on the combined
long- and short-exposure mosaics separately using the APEX output and
the ``aper'' IDL procedure from the IDLASTRO library. We used a 2
pixel aperture radius and a sky annulus of 2-6 pixels.  The
(multiplicative) aperture corrections we used follow the values given
in the IRAC Data Handbook: 1.213, 1.234, 1.379, and 1.584 for IRAC
channels 1, 2, 3, and 4, respectively. Flux densities have been
converted to magnitudes with the zero magnitude flux densities of
280.9 $\pm$ 4.1, 179.7 $\pm$ 2.6, 115.0 $\pm$ 1.7, and 64.1 $\pm$ 0.9
Jy for channels 1, 2, 3, and 4 respectively, as given in the IRAC Data
Handbook.   

To make a realistic estimate of the internal uncertainties, we
obtained photometry from the individual BCD images for the candidate
members.  We found that the scatter within that distribution is
generally of order 1\%.  Absolute uncertainties, arising from source
variability and uncertainties in the instrument calibration (primary
calibrators, aperture corrections, etc.), will be an additional error.
This absolute calibration uncertainty is estimated to be at least
$\sim$1.4\%. For the longer integration times, we estimated average
photometric errors in the four IRAC channels to be 0.02 mag for bright
stars, increasing to 0.1 mag  at 17.2, 16.6, 14.4, and 13.8 mag for
channels 1 to 4 respectively.   

The APEX source detection algorithm has a tendency to identify
multiple sources within the PSF of a single bright source, which can
cause significant confusion at the bandmerging stage. Because our flux
densities are determined from aperture photometry with a 2 pixel
aperture radius that is not deblended, any object with a companion
within 2 pixels is not actually resolved. The photometry lists were
therefore purged of multiple sources prior to bandmerging; objects
with an accidental companion within 2 pixels had the lower
signal-to-noise source removed. 

We extracted photometry from the long and short exposure mosaics
separately for each channel, and merged these source lists together by
position using a search radius of 2 pixels (2$^{\prime\prime}$.44) to
obtain a catalog for each channel. The magnitude cutoff transition,
where we used the long exposure rather than the short exposure
photometry, corresponds to magnitudes of 11, 10, 8.4, and 7.5 mag for
channels 1 through 4, respectively. We obtained $\sim$40,000 sources
in channel 1,  $\sim$34,000 sources in channel 2, $\sim$7500 in
channel 3, and $\sim$5000 sources in channel 4.  These relative
numbers are consistent with expectations from other star-forming
regions and with the sensitivities of the  channels.

\begin{deluxetable}{lccl}
\tablecaption{Summary of IRAC and MIPS observations\label{table:aors}}
\tablewidth{0pt}
\tablehead{
\colhead{Program}  &  \colhead{AORKEY} & \colhead{map center} &
\colhead {date obtained}}
\startdata
\cutinhead{IRAC} 
235 & 13375232 &  5:07:33.00 -6:17:00.0 & 2005-03-25\\ 
266 & 16940288 &  5:05:52.00 -6:27:52.0 & 2006-03-24\\ 
266 & 16941056 &  5:03:55.00 -6:42:22.0 & 2006-03-24\\ 
266 & 16941312 &  5:05:17.00 -7:10:38.0 & 2006-03-23\\ 
266 & 16941568 &  5:04:24.00 -7:41:48.0 & 2006-03-24\\ 
266 & 16941824 &  5:03:10.00 -7:04:57.0 & 2006-03-24\\ 
462 & 24248832 &  5:06:29.50 -6:52:58.1 & 2008-03-09\\ 
462 & 24249088 &  5:07:18.39 -6:07:11.1 & 2007-10-16\\ 
462 & 24249344 &  5:07:31.60 -6:21:50.1 & 2007-10-16\\ 
\cutinhead{MIPS} 
235 & 13374976 &  5:07:33.00 -6:17:00.0 & 2005-03-10\\
266 & 16940544 &  5:06:10.00 -6:47.47.0 & 2006-03-31\\
266 & 16940800 &  5:03:42.00 -7:29:19.0 & 2006-03-31\\
\enddata
\end{deluxetable}

\subsection{MIPS Data Reduction}

For MIPS, we started with the SSC-pipeline-produced BCDs, version
S16.1.  The 24 \mum\ data were processed differently
from the 70 and 160 \mum\ data.

For 24 \mum, we used MOPEX to construct a mosaic from the BCDs, with
the pixel scale set at 2.45$^{\prime\prime}$, very close to native
pixel size. We used the APEX 1-Frame routines and the point-response
function (PRF) provided on the SSC website to perform point-source
PRF-fitting photometry on the mosaics.  We determined via visual
inspection that 24 \mum\ sources with an APEX-extracted
signal-to-noise ratio (SNR) below 7 were not reliable; therefore, we
did not include those sources in the following analysis.  Moreover, we
inspected individual sources and removed \update{13}{13-Sept} more
clearly false detections (arising from nebulosity  or diffraction
patterns around bright stars).  This resulted in a catalog of
\update{1082}{13-Sept} point sources ranging from
\update{0.9~mJy}{13-Sept}\ to \update{700~mJy}{13-Sept}.  The zero
point used to convert flux densities to magnitudes is 7.14~Jy, based
on the extrapolation from the Vega spectrum as published in the MIPS
Data Handbook.  We took the error from the signal-to-noise ratio
returned by APEX as the baseline statistical error. However, the true
uncertainty is affected by flat fielding issues, etc., and is
estimated to be $\sim$4\%.  We added a 4\% flux error in quadrature to
the  APEX-derived error to estimate our final 24 $\mu$m uncertainties.

At 70 and 160 $\mu$m, the individual unfiltered frames (more than
5400 BCD files) were combined using MOPEX, setting their pixel scales
to their native values of 9.8$^{\prime\prime}$ and 16$^{\prime\prime}$
for 70 and 160 \mum, respectively. The final mosaic at 70 \mum\ has a
peak surface brightness of $\sim 100$ MJy~sr$^{-1}$, which is within
the linear  regime of the 70 \mum\ array \citep{Gordon-2007}; for the
160 \mum\ mosaic, the value is $\sim 200$ MJy~sr$^{-1}$ which is also
in the linear regime of the array \citep{Stansberry-2007}, and
therefore we expect at most a 20\% absolute flux density uncertainty.
The data were further processed to maintain the structure of the
extended emission and preserve its calibration.  For the 160 \mum\
mosaic, we applied a two-dimensional (3$\times$3 native pixel) median
boxcar filter to interpolate across  missing (``NaNs" or ``not a
number'') data due to the incomplete coverage (see above). This step
affects only those pixels near the ``NaNs" by a few percent. For both
Ge channels, we destriped the image using a ridgelet algorithm
(J.~Ingalls, private communication), which conserves flux and reduces
the noise. 

For the 70 \mum\ point sources, we performed point-source photometry
in the same fashion as we did for 24 \mum, using APEX 1-Frame on the
mosaic constructed as described, and the SSC-provided PRF.  Sources
with SNR below 8 were rejected, and we carefully removed obvious false
detections by individual inspection. This resulted in a catalog of
\update{84}{13-Sept}\ point sources. The zero point used for
conversion between flux densities and magnitudes for the 70 $\mu$m
data is 0.77~Jy, again from the MIPS Data Handbook.

No point sources were detected in the 160 \mum\ data.

\subsection{Optical Observations and Data Reduction}

We obtained $UVR_cI_c$ images by observing with the United States
Naval Observatory (USNO) 40-inch telescope (Flagstaff, AZ) over
several epochs between November 2006 and January 2007.  An area of 2.8
square degrees was covered in all 4 bands by mosaicking the
$\sim$23.9$^\prime\times$23.2$^\prime$ field of view. We acquired at
least two images at two different exposure times for each pointing;
see Table~\ref{tab:optical_obs}.  On 15 March 2010, an additional
epoch of relatively short ($\leq$60 sec) integrations was obtained in
$VR_cI_c$ just of the head of the nebula (where our YSO candidates are
located; see below) as a final check on our measurements. Aperture
photometry was performed on the objects in each frame, magnitudes
between $\sim$10-12 in the shallowest integrations and $\sim$20-22 in
each band in the deepest integrations, subject to variations in the
quality of each night. Absolute photometric calibration was achieved
through observation of Landolt standards during each night. 

The additional epoch in 2010 allowed us to check our calibration and
assess intrinsic variability in these stars on $\sim$3 year
timescales.  Several of the YSO candidates are variable, some highly
variable ($>$0.2 mag; see Table 4). The three most variable sources
are all embedded in reflection nebulae.



In order to include the optical data in our spectral energy
distributions, we needed to convert the measured magnitudes to flux
densities in Janskys. For the $U$ and $V$ filters, we used the Johnson
zero points from Allen's Astrophysical Quantities (Fourth Edition,
2001, Arthur N. Cox (ed.), Springer-Verlag), specifically 1823 and
3781 Jy, respectively.  For the $R_c$ and $I_c$ filters, we used the
zero points from \cite{Bessell-1979}, which are 3040 and 2433 Jy,
respectively.

\ptlandscape
\begin{deluxetable}{lccp{1.3cm}p{1.3cm}p{1.3cm}p{1.3cm}p{2.3cm}p{2.3cm}p{2.3cm}p{2.3cm}}
\rotate
\tablewidth{21.8cm}
\tabletypesize{\tiny}
\tablecaption{Summary of our $UVR_cI_c$ observations from
Nov.~2006--Jan.~2007\label{tab:optical_obs}}
\tablehead{\colhead{\#} & \colhead{RA (2000)} & \colhead{Dec
(2000)} & \colhead{$U$ exp.\ (s)} & \colhead{$V$ exp.\ (s)} & \colhead{$R_c$
exp.\ (s)} & \colhead{$I_c$ exp.\ (s)} & \colhead{$U$ date} & \colhead{$V$
date} & \colhead{$R_c$ date} & \colhead{$I_c$ date}}
\startdata
   1 & 05:08:14.820 & -6:16:39.062 &     1800, 180 &               900, 60 &                20, 600 &                         60, 600, 10 &           18/11/06, 17/11/06 &                               18/11/06, 17/11/06 &                                         17/11/06, 18/11/06 &                                                             17/11/06, 18/11/06, 17/11/06\\ 
   2 & 05:08:12.326 & -6:16:49.188 & 900, 180, 360 & 900, 180, 60, 120, 30 & 5, 900, 180, 20, 30, 2 & 20, 30, 2, 5, 900, 180, 60, 600, 10 & 31/10/06, 17/11/06, 09/01/07 & 31/10/06, 31/10/06, 20/11/06, 09/01/07, 21/11/06 & 31/10/06, 31/10/06, 31/10/06, 20/11/06, 31/10/06, 18/11/06 & 21/11/06, 31/10/06, 18/11/06, 31/10/06, 31/10/06, 31/10/06, 20/11/06, 18/11/06, 21/11/06\\ 
   3 & 05:08:11.814 & -6:04:28.112 &     1800, 180 &               900, 60 &                20, 600 &                         60, 600, 10 &           18/11/06, 17/11/06 &                               18/11/06, 17/11/06 &                                         17/11/06, 18/11/06 &                                                             17/11/06, 18/11/06, 17/11/06\\ 
   4 & 05:06:54.960 & -6:13:32.816 & 180, 600, 360 &          900, 60, 120 &                20, 600 &                         60, 600, 10 & 17/11/06, 19/11/06, 09/01/07 &                     19/11/06, 17/11/06, 09/01/07 &                                         17/11/06, 19/11/06 &                                                             20/11/06, 19/11/06, 19/11/06\\ 
   5 & 05:05:35.082 & -6:13:31.057 &      180, 600 &               900, 60 &                20, 600 &                         60, 600, 10 &           17/11/06, 19/11/06 &                               19/11/06, 17/11/06 &                                         17/11/06, 19/11/06 &                                                             17/11/06, 19/11/06, 19/11/06\\ 
   6 & 05:06:54.658 & -6:33:28.589 &      180, 600 &               900, 60 &                20, 600 &                         60, 600, 10 &           17/11/06, 20/11/06 &                               19/11/06, 17/11/06 &                                         17/11/06, 19/11/06 &                                                             19/11/06, 19/11/06, 19/11/06\\ 
   7 & 05:05:34.572 & -6:33:29.144 &      180, 600 &               900, 60 &            20, 600, 40 &                     20, 60, 600, 10 &           17/11/06, 21/11/06 &                               21/11/06, 17/11/06 &                               17/11/06, 21/11/06, 09/01/07 &                                                   09/01/07, 17/11/06, 21/11/06, 17/11/06\\ 
   8 & 05:04:14.580 & -6:33:27.864 &      180, 600 &               900, 60 &            20, 60, 600 &                         60, 600, 10 &           17/11/06, 21/11/06 &                               21/11/06, 17/11/06 &                               17/11/06, 21/11/06, 21/11/06 &                                                             21/11/06, 21/11/06, 17/11/06\\ 
   9 & 05:07:25.289 & -6:48:31.295 &      180, 600 &               900, 60 &                20, 600 &                      60, 600, 10, 2 &           17/11/06, 21/11/06 &                               21/11/06, 17/11/06 &                                         17/11/06, 21/11/06 &                                                   21/11/06, 21/11/06, 17/11/06, 21/11/06\\ 
  10 & 05:06:05.416 & -6:48:28.150 & 180, 600, 360 &          900, 60, 120 &           900, 20, 600 &                15, 900, 60, 600, 10 & 17/11/06, 21/11/06, 09/01/07 &                     21/11/06, 17/11/06, 09/01/07 &                               09/01/07, 17/11/06, 19/01/07 &                                         09/01/07, 09/01/07, 17/11/06, 19/01/07, 17/11/06\\ 
  11 & 05:04:47.396 & -6:48:48.174 &      180, 600 &      900, 60, 600, 10 &        900, 20, 600, 2 &                 900, 60, 600, 10, 2 &           17/11/06, 19/01/07 &           19/01/07, 17/11/06, 09/01/07, 09/01/07 &                     09/01/07, 17/11/06, 19/01/07, 17/11/06 &                                         09/01/07, 17/11/06, 19/01/07, 17/11/06, 17/11/06\\ 
  12 & 05:03:25.438 & -6:48:44.163 &      180, 600 &          900, 60, 600 &           900, 20, 600 &                    900, 60, 600, 10 &           20/11/06, 19/01/07 &                     19/01/07, 20/11/06, 09/01/07 &                               09/01/07, 20/11/06, 19/01/07 &                                                   09/01/07, 20/11/06, 19/01/07, 20/11/06\\ 
  13 & 05:07:24.998 & -7:08:25.470 &     1800, 180 &               900, 60 &                20, 600 &                         60, 600, 10 &           20/11/06, 20/11/06 &                               20/11/06, 20/11/06 &                                         20/11/06, 20/11/06 &                                                             20/11/06, 20/11/06, 20/11/06\\ 
  14 & 05:06:07.763 & -7:08:48.399 &      180, 600 &           900, 60, 10 &            20, 600, 10 &                     15, 60, 600, 10 &           20/11/06, 20/03/07 &                     20/03/07, 20/11/06, 20/11/06 &                               20/11/06, 20/03/07, 20/11/06 &                                                   15/01/07, 20/11/06, 20/03/07, 20/11/06\\ 
  15 & 05:04:48.979 & -7:09:27.985 &      180, 600 &       900, 60, 30, 10 &             20, 600, 2 &                      60, 600, 10, 2 &           20/11/06, 15/02/07 &           15/02/07, 20/11/06, 20/11/06, 20/11/06 &                               20/11/06, 15/02/07, 20/11/06 &                                                   20/11/06, 15/02/07, 20/11/06, 20/11/06\\ 
  16 & 05:03:24.572 & -7:08:17.494 &      180, 600 &        3, 900, 60, 30 &             20, 600, 2 &                      60, 600, 10, 2 &           20/11/06, 15/02/07 &           20/11/06, 15/02/07, 20/11/06, 20/11/06 &                               20/11/06, 15/02/07, 20/11/06 &                                                   20/11/06, 15/02/07, 20/11/06, 20/11/06\\ 
  17 & 05:07:18.477 & -7:28:30.082 &           600 &           900, 60, 30 &            20, 60, 600 &                         60, 600, 10 &                     18/03/07 &                     18/03/07, 20/11/06, 20/11/06 &                               20/11/06, 20/11/06, 18/03/07 &                                                             20/11/06, 18/03/07, 20/11/06\\ 
  18 & 05:06:00.376 & -7:28:55.094 &      180, 600 &           900, 60, 30 &            20, 600, 10 &                      60, 600, 10, 2 &           20/11/06, 19/03/07 &                     19/03/07, 20/11/06, 20/11/06 &                               20/11/06, 19/03/07, 20/11/06 &                                                   20/11/06, 19/03/07, 20/11/06, 20/11/06\\ 
  19 & 05:04:36.463 & -7:29:11.645 &      180, 600 &          900, 60, 600 &                20, 600 &                      60, 600, 10, 2 &           20/11/06, 17/03/07 &                     17/03/07, 20/11/06, 17/01/07 &                                         20/11/06, 17/03/07 &                                                   20/11/06, 17/03/07, 20/11/06, 20/11/06\\ 
  20 & 05:05:57.436 & -7:48:33.482 &      180, 600 &           900, 60, 30 &            20, 600, 10 &                      60, 600, 10, 2 &           20/11/06, 18/01/07 &                     18/01/07, 20/11/06, 20/11/06 &                               20/11/06, 18/01/07, 20/11/06 &                                                   20/11/06, 18/01/07, 20/11/06, 20/11/06\\ 
  21 & 05:04:37.303 & -7:48:37.014 &           600 &           900, 60, 30 &                20, 600 &                         60, 600, 10 &                     18/01/07 &                     18/01/07, 20/11/06, 20/11/06 &                                         20/11/06, 18/01/07 &                                                             20/11/06, 18/01/07, 20/11/06\\ 
  22 & 05:05:55.785 & -8:08:07.414 &      180, 600 &           900, 60, 10 &             20, 600, 2 &                   5, 60, 600, 10, 2 &           21/11/06, 18/01/07 &                     18/01/07, 21/11/06, 21/11/06 &                               21/11/06, 18/01/07, 21/11/06 &                                         21/11/06, 21/11/06, 18/01/07, 21/11/06, 21/11/06\\ 
  23 & 05:04:32.104 & -8:09:00.677 &      180, 600 &           900, 60, 10 &             20, 600, 2 &                      60, 600, 10, 2 &           21/11/06, 18/01/07 &                     18/01/07, 21/11/06, 21/11/06 &                               21/11/06, 18/01/07, 21/11/06 &                                                   21/11/06, 18/01/07, 21/11/06, 21/11/06\\ 
\enddata
\end{deluxetable}

\subsection{Bandmerging}

We matched IRAC sources from adjacent channels to empirically
estimate our astrometric uncertainty. The histograms of coordinate
differences show a 1$\sigma$ uncertainty of $\sim$0$^{\prime\prime}$.3
in both directions. At this Galactic latitude and longitude
(207.4,$-$27.9),   the density of sources is relatively low ($\sim$10
detections per square arcmin in the 3.6 $\mu$m images, 4 detections
per square arcmin in $R_c$ and $I_c$ images). Via empirical tests
across all bands, we determined that a 2$\arcsec$ radius (1.6 IRAC
pixels) for source matching works for our sources across the optical,
2MASS, IRAC, and MIPS bands. For a matching radius less than
$\sim$1$^{\prime\prime}$, not all of the real matches are found. The
number of matches increases steeply with radius until 
$1^{\prime\prime}$; after that, the number of matches increases slowly
with radius until $\sim$4$^{\prime\prime}$ where the number of matches
is dominated by false source associations.  Using a matching
radius of 2$^{\prime\prime}$ appears to provide the best possible
merging of sources across bands, even for 24 \mum; the centroiding for
the 24 \mum\ data is much better than 2$^{\prime\prime}$, even though
the data have a spatial resolution of $\sim$6$^{\prime\prime}$. 

Because the 3.6 \mum\ observations are the most sensitive of our
Spitzer observations, we started by constructing a catalog based on
these detections. To be included this catalog, sources have to be
detected in 3.6 $\mu$m and in at least one of the other bands.  Out of
the \update{39,817}{16-Sept} sources in this catalog, \update{3365
(8.5\%)}{16-Sept} have been detected in all 4 IRAC channels, 15\% in
the $JHK_s$  bands, \update{50\%}{16-Sept} in the $I_c$ band (recall
that the optical observations are deeper than the 2MASS
observations),  and just \update{12\%}{16-Sept} in the $U$ band.  At
24 \mum, 1\% (426 sources) of the 3.6 \mum-based catalog are
detected.  However, since the spatial coverage provided by our MIPS
map is larger than that for IRAC, and since some very embedded young
objects may not be detected in shorter wavelengths than 24 $\mu$m, we
have also included in the catalog those sources with 24 $\mu$m
detections even if they do not have 3.6 $\mu$m detections. Out of the
1082 sources detected in 24 $\mu$m, \update{48\%}{16-Sept} are
detected in the $I_c$ band and \update{27\%}{16-Sept} in the $U$
band.  However, the region mapped by MIPS is not exactly matched to
the optical observations; considering just the region covered by both
the optical and MIPS maps, these numbers become \update{69\%}{16-Sept}
and \update{39\%}{16-Sept}, respectively. Similarly,
\update{30\%}{16-Sept} of the 24 $\mu$m sources are detected in all 4
IRAC bands; if one is restricted to the area covered by both MIPS and
IRAC maps, 70\% are detected at all 4 IRAC bands plus MIPS-24.
Finally, \update{38\%}{16-Sept} of the 24 $\mu$m sources are detected
in $JHK_s$.

\section{Selection of YSO candidates with IR excess}
\label{sec:findthem}

With our new multi-wavelength view of the molecular cloud, we can begin
to look for the members of this complex. We focus on finding sources
having an infrared excess characteristic of YSOs surrounded by a dusty
disk.  We do not expect to find a dense population of YSOs (see
\S\ref{sec:intro}).  Additionally, the molecular cloud is not  very
dense, so it will not obscure many of the background sources. 
Thus, we expect a high source contamination rate.  Separating the YSO
candidates from the extraneous sources requires an extensive
weeding-out of the galactic and extragalactic contaminants. To begin
this task, we identify objects with excesses at IRAC and MIPS
wavelengths.  Then, we use filtering mechanisms from the literature
based on Spitzer colors to distinguish likely galaxies from likely
members.  We also use the additional information we have from the
optical to further winnow the candidate list.


\subsection{IRAC selection}

\begin{figure}
\includegraphics[width=\hsize]{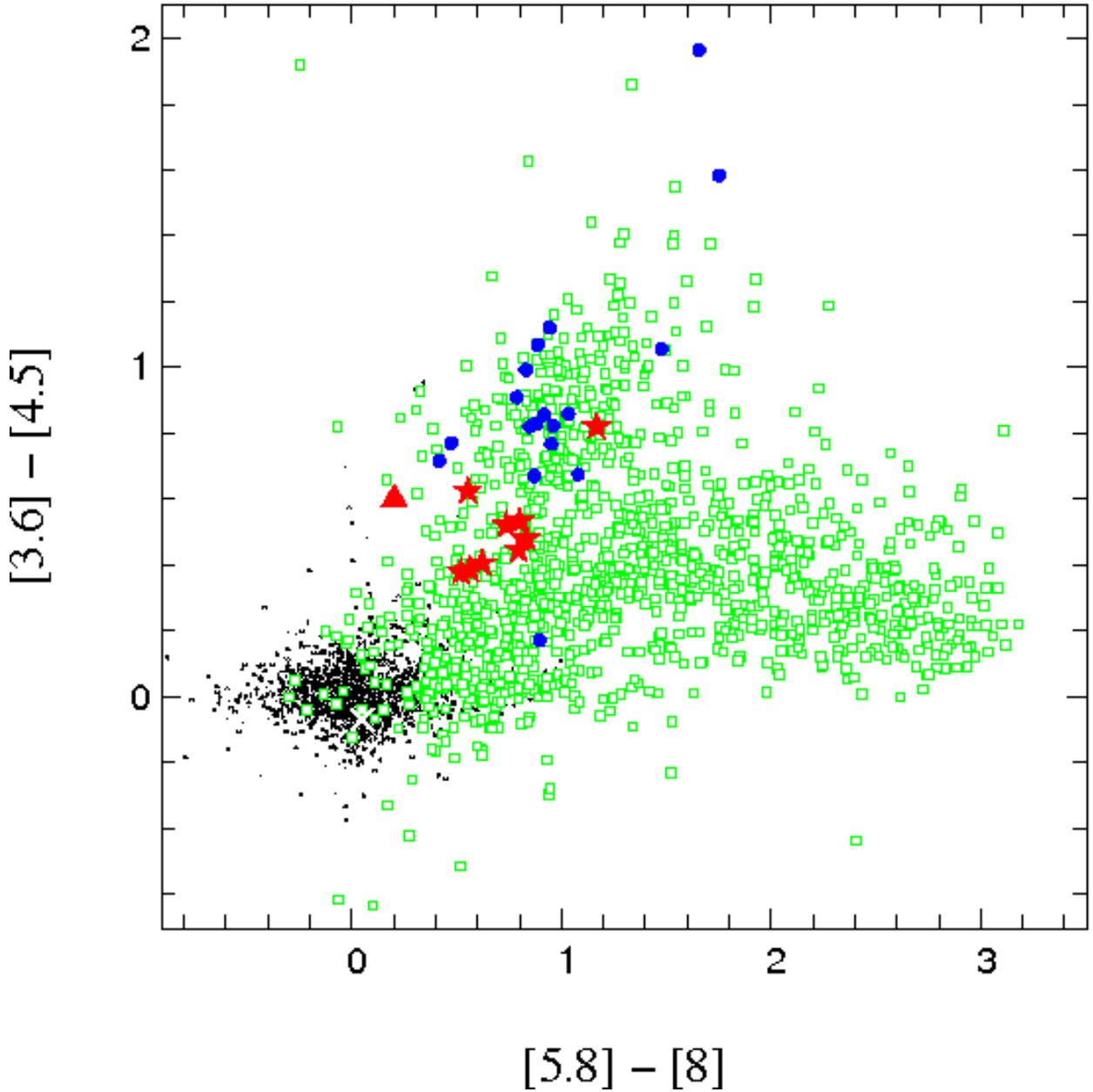}
\caption{The $[3.6]-[4.5]$ {\it versus} $[5.8]-[8]$ color-color
diagram.  Sources flagged as PAH emission sources (background
galaxies) or AGN by our selection method are plotted in green squares.
Red star symbols are selected YSOs candidates with $[3.6]<12$ and blue
points are YSO candidates selected in the same fashion but with
$[3.6]>12$.  The red triangle is an edge-on disk candidate (see
\S\ref{sec:mipssel}).  The black points are the remaining detections. 
The $\times$ symbol near zero color is the weak-lined  T~Tauri (wTTS)
candidate (2MASS~J05060574-0646151, see \S\ref{sec:prev}). See text
for more discussion of each of these categories.
\label{fig:i1i2i3i4}
}
\end{figure}

We first select sources using a set of IRAC color-magnitude and
color-color diagrams. Our selection is based on the
\cite{Gutermuth-2008a} method and is described in \cite{Guieu-2009};
in summary, this method uses cuts in several IRAC color-color and
color-magnitude spaces to identify likely YSOs and likely contaminants
such as active galactic nuclei (AGN). Using this technique, we
identified \update{459}{16-Sept} sources having colors consistent with
galaxies dominated by polycyclic aromatic hydrocarbon (PAH) emission
and \update{765}{16-Sept} more sources having colors consistent with
AGN. Just \update{27}{5-june} sources are not flagged as background
contaminants and have colors compatible with YSOs with an IRAC excess
in the $[3.6]-[5.8]/[4.5]-[8]$ plane (where the bracket notation
denotes magnitudes, e.g., [3.6] means magnitude at 3.6 $\mu$m).  
Figure~\ref{fig:i1i2i3i4} shows the IRAC color-color diagram for the
sources that passed and failed these color tests. (We note for
completeness that \cite{Gutermuth-2009} have updated their YSO
selection method with new rejection criteria; we have verified that
these new criteria do not affect our list of YSO candidates.)

Since we expect to find a lot of contamination here, the list of 27
sources not flagged as contaminants bears further scrutiny. These 27
sources consist of 9 sources brighter than [3.6]=12, with the
remaining 18 all fainter than this limit. The SWIRE (Spitzer Wide-area
Infrared Extragalactic Survey; \citealt{Lonsdale-2003}) observations
of the ELAIS N1 extragalactic field\footnote{VizieR Online Data
Catalog, II/255 \citep{Surace-2004}} consists of integrations of
similar depth, but this sample is expected to be essentially entirely
galaxies and foreground stars, and as such provides a guide to the
range of values expected from these contaminants.  The SWIRE survey
shows very few galaxies have [3.6] brighter than 12 magnitudes. The
IC~2118 objects that we have identified explicitly as contaminants
have 3.6 \mum\ flux densities comparable to the SWIRE contaminants,
with $[3.6]$ ranging from 12 to 18 mags.  But, out of the YSO
candidates, \update{18}{16 Sept} also have magnitudes fainter than
$[3.6] = 12$.  As the brightness decreases, the chances of these
objects being background contaminants increases.

\begin{figure}
\includegraphics[width=\hsize]{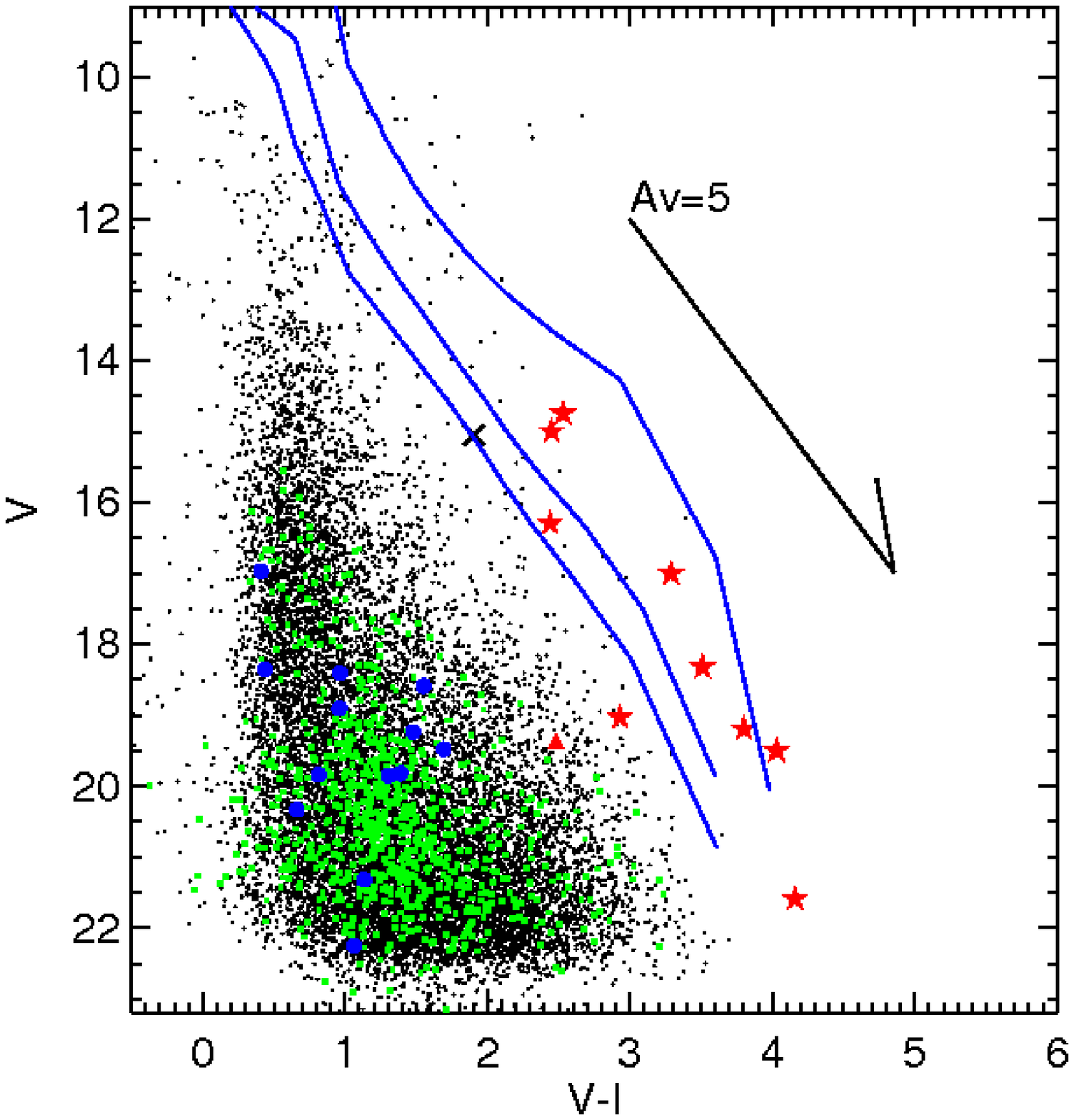}
\caption{
Optical $V$ {\it versus} $V-I_c$ color-magnitude diagram. The
  symbols are the same as in Figure \ref{fig:i1i2i3i4}.
  Note that the
  blue dots representing the [3.6]$>$12 sources are in fact plotted,
  but they are in amongst the green squares and black dots well below
  the ZAMS; this we take to be further evidence that the objects
  represented by the blue dots are contaminants and not YSO
  candidates. The red triangle is the edge-on disk candidate, which
appears here below the ZAMS, as appropriate for such an object. The
solid blue lines are models from Siess et al.~(2000) at 1, 10 and
30~Myr and scaled to 210~pc, where we have tuned the color-effective
temperature relation such that the 100 Myr isochrone matches that of
the Pleiades single-star sequence \citep{Stauffer-2007,Jeffries-2007}.
A reddening vector is also indicated.
  \label{fig:vvi}
}
\end{figure}

Since we have a wealth of supporting data, we can use these data to
additionally constrain our YSO selection.   Figure \ref{fig:vvi} shows
an optical color-magnitude diagram for these sources.   The optical
colors and magnitudes for these \update{18}{16 Sept} faint sources all
fall in the region occupied by main-sequence stars or background
galaxies. In contrast, the 9 remaining YSO candidates with $[3.6]<12$
all appear redder, near or above  a 30~Myr Siess et al.~(2000)
isochrone in the $V / V-I_c$ color magnitude diagram. We strongly
suspect that the faint ones are contaminants. The 9 brighter objects
selected in $[3.6]-[5.8]/[4.5]-[8]$ with $[3.6]<12$ are included in
our list of IRAC-selected YSO candidates. For reference, the
\update{18}{16-Sept} objects with  $[3.6]>12$ are represented with a
blue filled circle in most of the remaining figures in this paper. 
The restriction we have imposed that $[3.6]<12$ means that we are
effectively senstitive to masses greater than 0.1 $M_{\odot}$ at a
distance of 210 pc for an assumed age of $\sim$3-5 Myr.

All of these 9 IRAC-selected YSO candidates are seen at MIPS-24.  This
is in contrast to the catalog as a whole, where only 10\% of the
sources seen in all four IRAC bands are also seen at 24 \mum.  Since
legitimate members of IC~2118 with infrared excesses located at
$\sim$210 parsecs should be detected at 24 \mum\ by our observations,
the fact that we see these objects gives us further confidence that
these objects are truly YSOs.  Three of them are also seen at 70 \mum;
similarly, legitimate YSOs with large excesses should be detected by
our shallow 70 \mum\ observations, so this also suggests that these
objects are true YSOs.  (Note, of course, that most of the sources
seen in our observations at 24 or 70 \mum\ are foreground stars or
background galaxies, so detection at any MIPS band by itself is
insufficient evidence for youth.)

We extracted and examined the POSS, 2MASS, and our optical images for
each of these candidates to verify that they did not appear extended
in any of these bands\footnote{Some of our high school students,
trained in observational techniques using the Sloan Digital Sky Survey
Galaxy Zoo, participated in this classification.  At least three
different individuals examined images of each of more than 400
candidates selected by a variety of Spitzer color selection mechanisms
in order to identify likely galaxies.}.  While all extended objects
may not be galaxies (many YSOs in Taurus at 140 pc appear as extended
objects due to circumstellar reflection nebulae, edge-on disks, jets,
etc.; see \citealt{Rebull-2009}), extended objects seen here are much
more likely to be galaxies than YSOs at 210 pc.  All of our YSO
candidates passed this check and appear to be point sources in all
available bands.

\subsection{MIPS selection}
\label{sec:mipssel}

\begin{figure}
\includegraphics[width=\hsize]{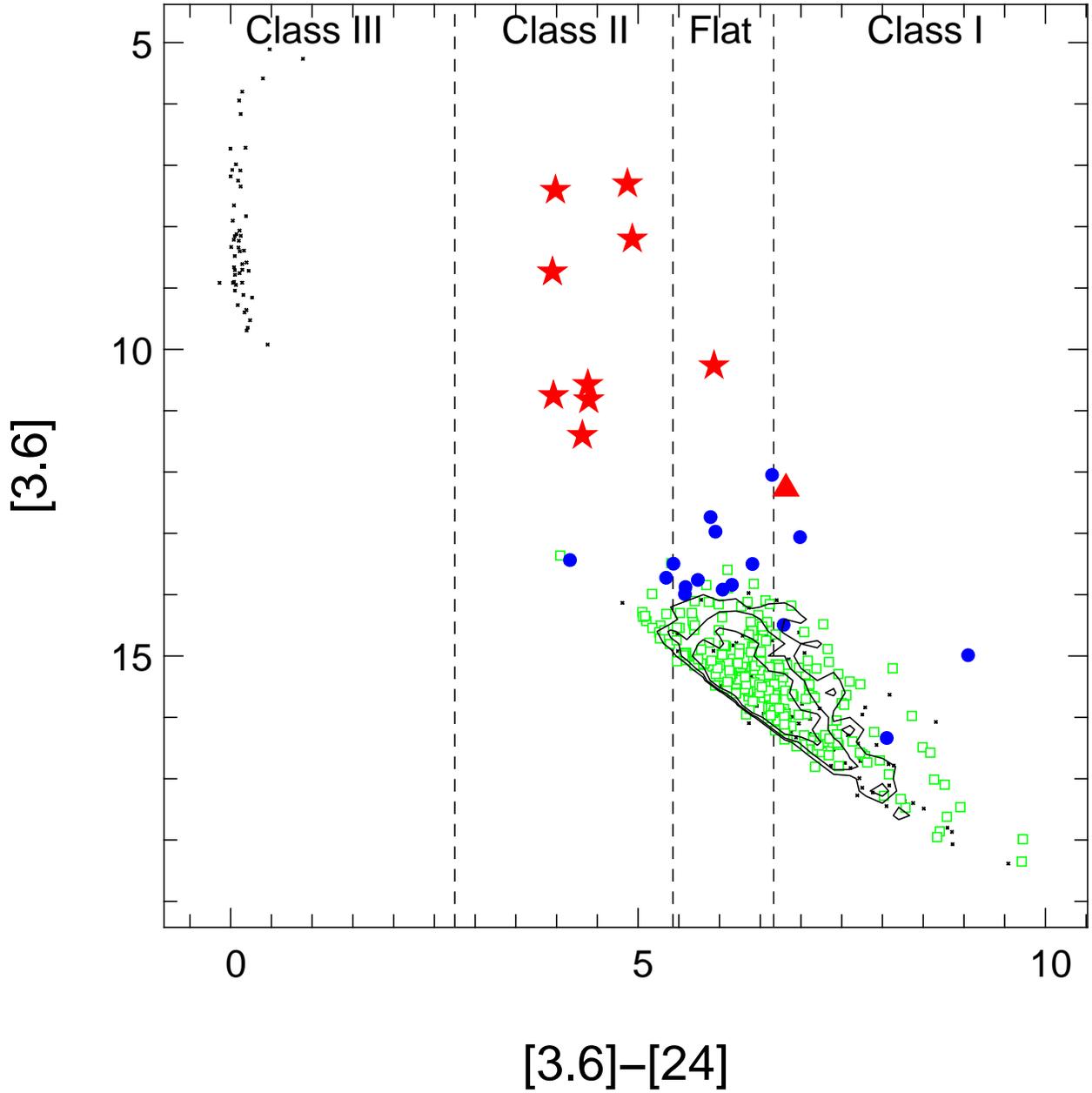}
\caption{ The [3.6]/ [3.6]-[24] color magnitude diagram. Symbols are
  the same as in Figure \ref{fig:i1i2i3i4}.  The  black curves are
  contours from the SWIRE catalog truncated to our IC~2118 detection
  limit ({\it i.e.,} $[24]<9.4$); see text.  \label{fig:i1i1m1}}
\end{figure}

This IRAC-based selection method does not allow us to detect young
stars having an infrared excess only at wavelengths longer than
8~\mum.  We looked for potential sources with a $[3.6]-[24]$ color
excess but no IRAC excess, which would be intermediate between
photospheric objects and Class~II stars; these objects are Class~III
sources with weak excesses (possible transitional disks). 
Figure~\ref{fig:i1i1m1} shows the $[3.6] / [3.6]-[24]$ color-magnitude
diagram. In this parameter space, photospheres have zero color, and
galaxies are red and faint.  This figure also contains data from the
SWIRE ELAIS N1 field.  We have truncated the SWIRE catalog to fit our
shallower IC~2118 24~$\mu$m detection limit, rejecting SWIRE sources
with [24]$>$9.4, and plotted the catalog here as black contours. The
SWIRE sample is essentially entirely galaxies and foreground stars,
and as such provides a visual guide to the locations where such
objects appear in the diagram. 

The most striking thing about this diagram for IC~2118 is that there
are three distinct groups of objects: objects of zero color (likely
foreground/background stars), objects that are faint and red (likely
galaxies), and objects that are bright and red (likely YSOs).  No
sources appear between the photospheric/Class III objects and the
Class~II objects. The bright objects we selected above based on IRAC
colors are also bright here, strongly suggesting the presence of an
infrared excess we interpret as circumstellar dust.  The contaminants
identified above are also identified as likely contaminants here;
these objects nearly all fall on top of the SWIRE contours. The
objects identified as having YSO-like IRAC colors but being as faint
as the contaminants are also faint and have contaminant-like colors in
this parameter space.  Within the faint contaminant portion of the
diagram, the YSO candidates are among the brightest or reddest
objects, which is a likely result of our selection process.  This
lends further support to our assertion that the faint objects are
likely background galaxies.

Despite the fact that Figure~\ref{fig:i1i1m1} does not reveal any new
candidate YSOs near the IRAC-identified candidate YSOs, it does
identify another source of particular interest.  Figure
\ref{fig:i1i1m1} highlights a source (as a red triangle) that is red,
with a $[3.6]-[24]$ color of $\sim$6.5 and a moderately faint
[3.6]$\sim$12, which is among the brighter values for the likely
contaminants.  This source (named 050721.9-061152) has not strictly
been selected as a YSO candidate or background contaminant by our
IRAC-based method, but it is located in the border of our YSO
selection area in the $[3.6]-[5.8]/[4.5]-[8]$ diagram; it has a clear 
$[3.6]-[5.8]$ excess but a low $[4.5]-[8]$ excess. Somewhat
surprisingly, it is clearly detected at 70 $\mu$m.  Since our 70 \mum\
survey is shallow enough that we would only detect IC~2118 YSOs if
they have a large infrared excess, we take that detection, plus the
marginal shorter-band color selections, as evidence that this object may
very well be a YSO candidate, and we add it to our list.  We discuss
this object again below.

We use [3.6] vs.\ $[3.6]-[24]$ to look for YSOs, whereas many papers
in the literature (e.g., \citealt{Rebull-2009} and references therein)
use the $K_s$ vs.\ $K_s-[24]$ color-magnitude space to find YSOs.  In
that diagram, the $K_s-[24]$ colors for M stars deviate significantly
from zero \citep{Gautier-2007}, so objects without disks but with a
(possibly unknown) M spectral type will appear to have an infrared
excess. The color departure from 0 for M stars using $[3.6]-[24]$
happens at later types than for $K_s-[24]$, which makes it a better
choice to use for searches for small excesses in cases where the
spectral type of the candidates is unknown.  However, one can
obviously only construct a [3.6] vs.\ $[3.6]-[24]$ diagram for those
regions where one has both IRAC and MIPS data. Since our maps do not
match exactly, and since there are previously-identified T~Tauris well
off of the densest parts of the nebula, we also constructed a $K_s$
vs.\ $K_s-[24]$ color-magnitude diagram using 2MASS and MIPS data to
see if any YSO candidates appeared off the edge of our IRAC maps.  Two
new sources appear as potential YSOs here, but by inspection of the
images, both of these sources are extended, and one (2MASX
J05015386-0745321) is a known galaxy.  We do not include either of
these objects on our YSO candidate list. 

One previously-identified T~Tauri star (RXJ~0502.4-0744) is located in
our MIPS map but not in the IRAC map, and is detected at 24 $\mu$m;
however this object does not show $K_s-[24]$ infrared excess. Its
measured flux density is 1.3$\pm$0.12\,mJy at 24 $\mu$m
([24]=9.36$\pm$0.1).

\section{Properties of the YSO candidates}
\label{sec:properties}

\subsection{Magnitudes and Spectral Energy Distributions}

\begin{figure}
\plotone{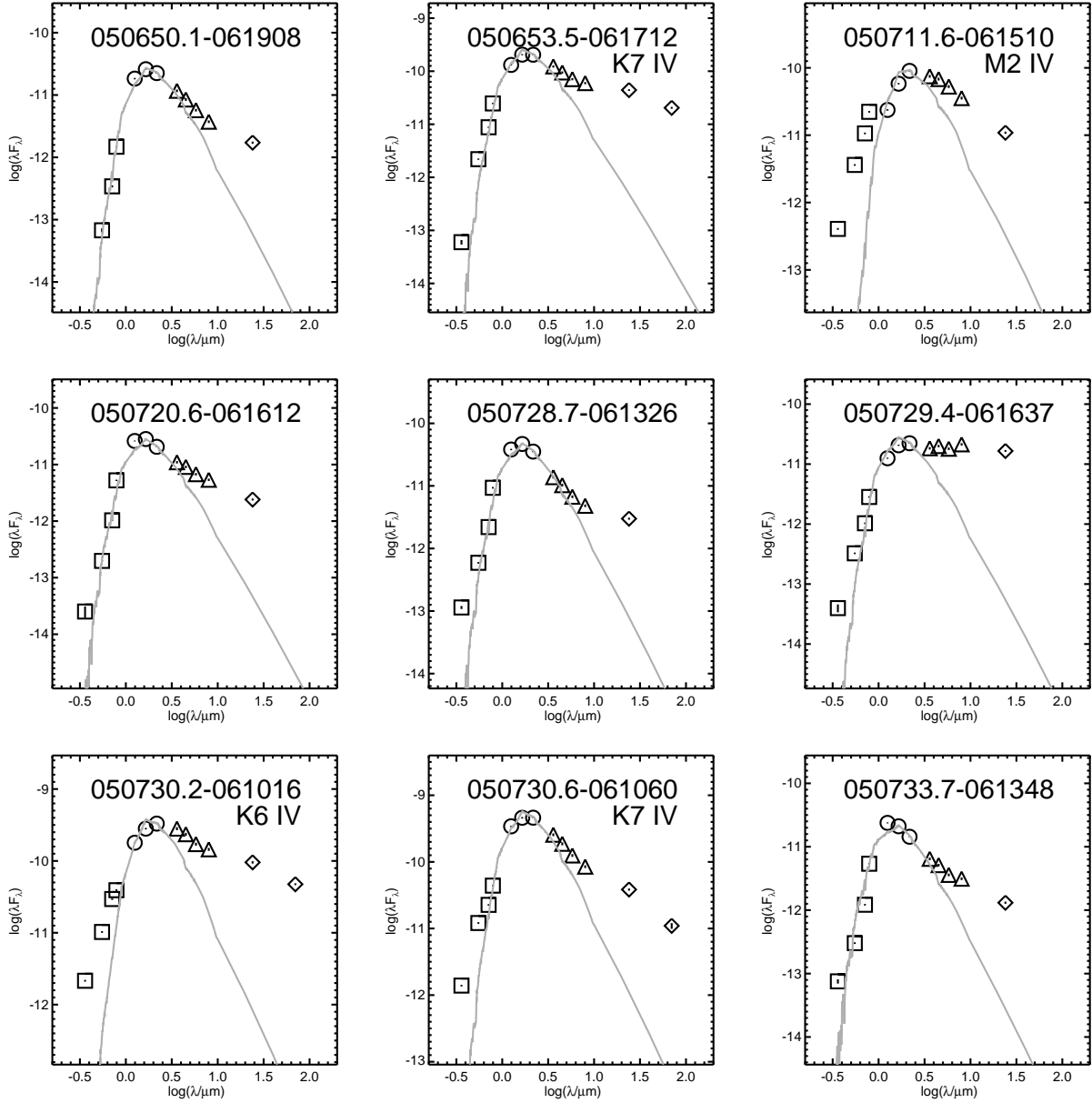}
\caption{SEDs for 9 of our YSO candidates; the 10th is in
Figure~\ref{fig:SED70}.  The squares are our optical points, the
circles are 2MASS, the triangles are IRAC, and the diamonds are MIPS.
Error bars are indicated, and in most cases are smaller than the
symbol. Spectral types are known for some objects from \cite{Kun-2004}, 
and those are indicated where possible.  In cases where no
type is known, an M0 is assumed. Reddened models from the
Kurucz-Lejeune model grid \citep{Lejeune-1997,Lejeune-1998} are shown for
reference; see text. \label{fig:seds}}
\end{figure}

Coordinates and our measured magnitudes between $U$ and 70 \mum\ for
our 10 YSO candidates appear in Table~\ref{tab:yso}.  Errors in the
optical photometry for these 10 objects are generally dominated by the
systematic errors, and as such are essentially all
$\sim$0.1 mag at $U$, 0.03-0.05 mag at $V$, $R_c$, and $I_c$,
0.02-0.03 mag at $JHK_s$ (where these values come from 2MASS), and
0.02 mag at all four IRAC bands. The 24 $\mu$m errors are typically
0.04 mag; the 70 $\mu$m errors vary enough between objects that they
appear separately in the table.   Spectral energy distributions (SEDs)
for the 9 IRAC-selected YSO candidates appear in
Figure~\ref{fig:seds}.   To guide the eye, we wished to add reddened
stellar models to these plots, but spectral types are only known for
four of the nine candidates.  In order to provide a reference, for the
remaining objects, we assumed an M0 type.  Thus, for each object in
Figure~\ref{fig:seds}, a reddened model is shown, selected from the
Kurucz-Lejeune model grid  \citep{Lejeune-1997,Lejeune-1998} and
normalized to $H$ band.  Note that this is not meant to be a robust
fit to the object, but rather a representative stellar SED to guide
the eye such that infrared excesses are immediately apparent.

\begin{figure}
\includegraphics[width=\hsize]{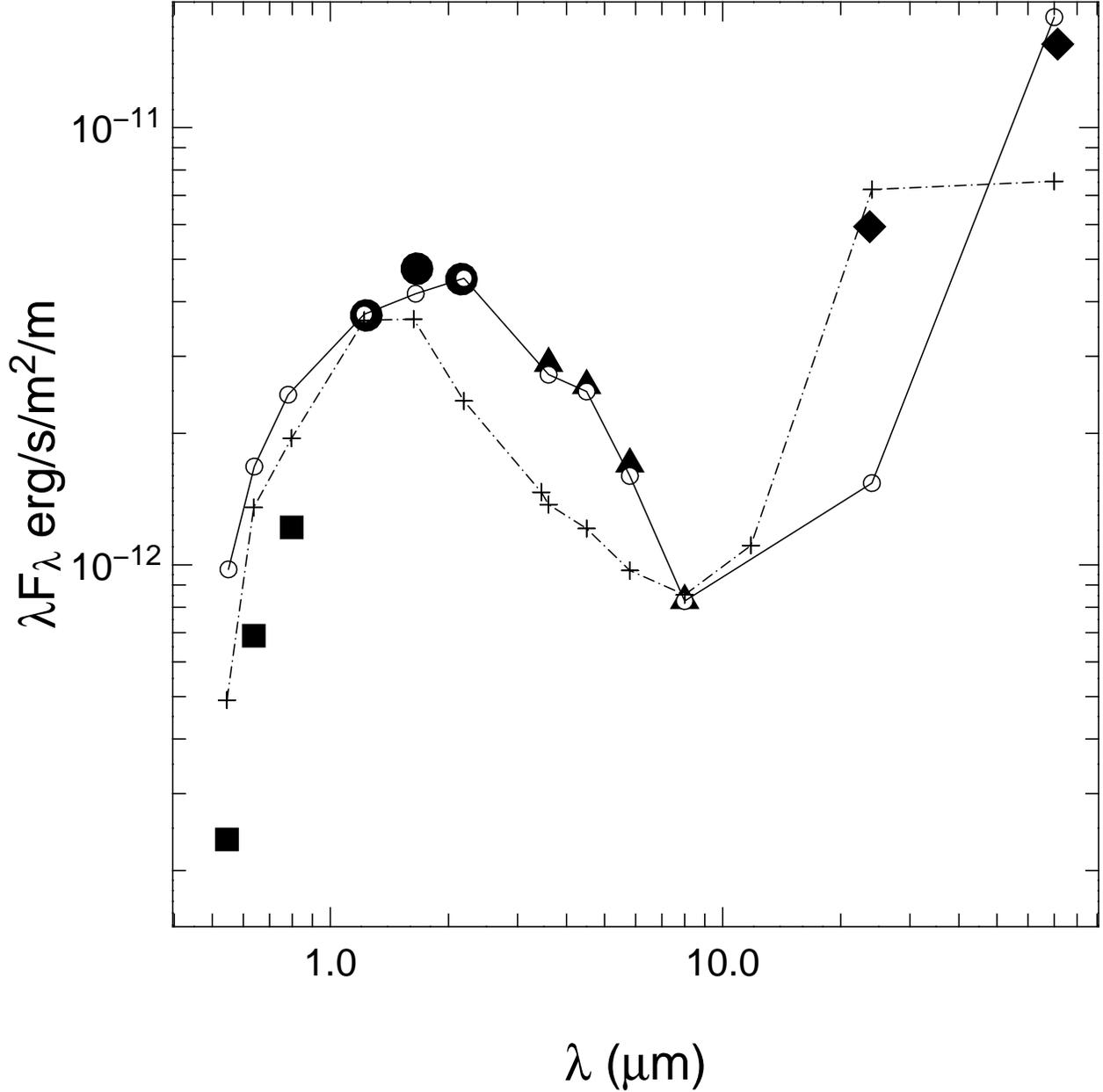}
\caption{SED of 050721.9-061152 (black filled symbols), compared to the
  well known edge-on disks HH~30 (solid line, circles; \citealt{Burrows-1996}) and
HV Tau C (dashed line, plus symbol; \citealt{Duchene-2010}).  These SEDs 
 have been normalized to 050721.9-061152 in
  the J band.\label{fig:SED70} 
  For the black filled points, the symbols are as in the previous
  Figure -- squares are our optical points, the
  circles are 2MASS, the triangles are IRAC, the diamonds are
  MIPS, and the error bars are smaller than the points.}
\end{figure}

\begin{figure}
\includegraphics[width=\hsize]{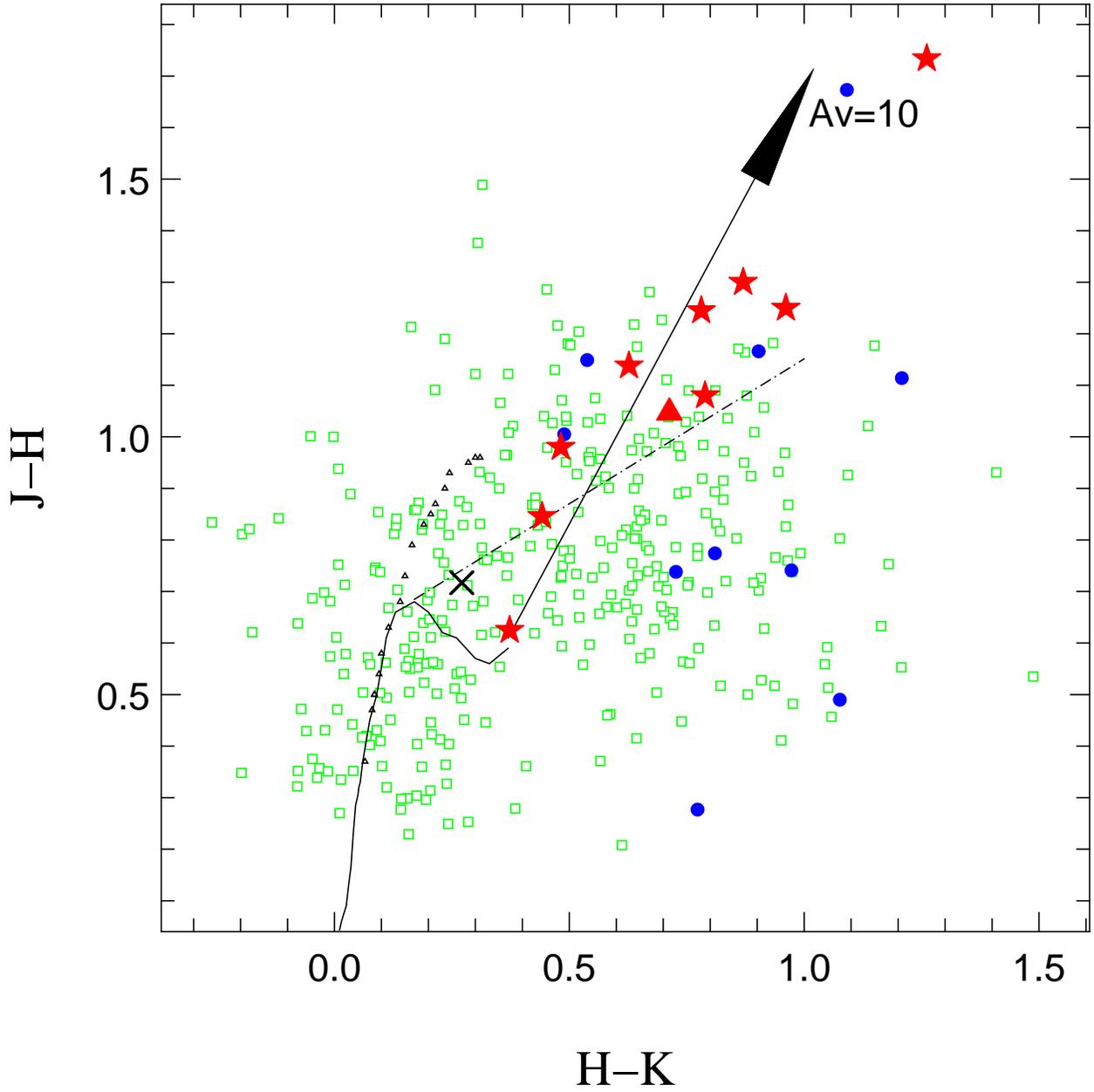}
\caption{ $J-H$ versus $H-K_s$ diagram for the sample, with the same
notation as earlier figures.   The T~Tauri locus is also indicated
(dash-dot line), as is a reddening vector (the arrow). The main
sequence is drawn by a solid line and the giant branch by small black
triangles. The $\times$ symbol is the wTTS candidate
(2MASS~J05060574-0646151, see \S\ref{sec:prev}).    This diagram shows
that all but one of the YSOs have an infrared excess starting at
H-band and that they have moderate reddening.  \label{fig:jhhk}}
\end{figure}

\begin{figure}
\includegraphics[width=\hsize]{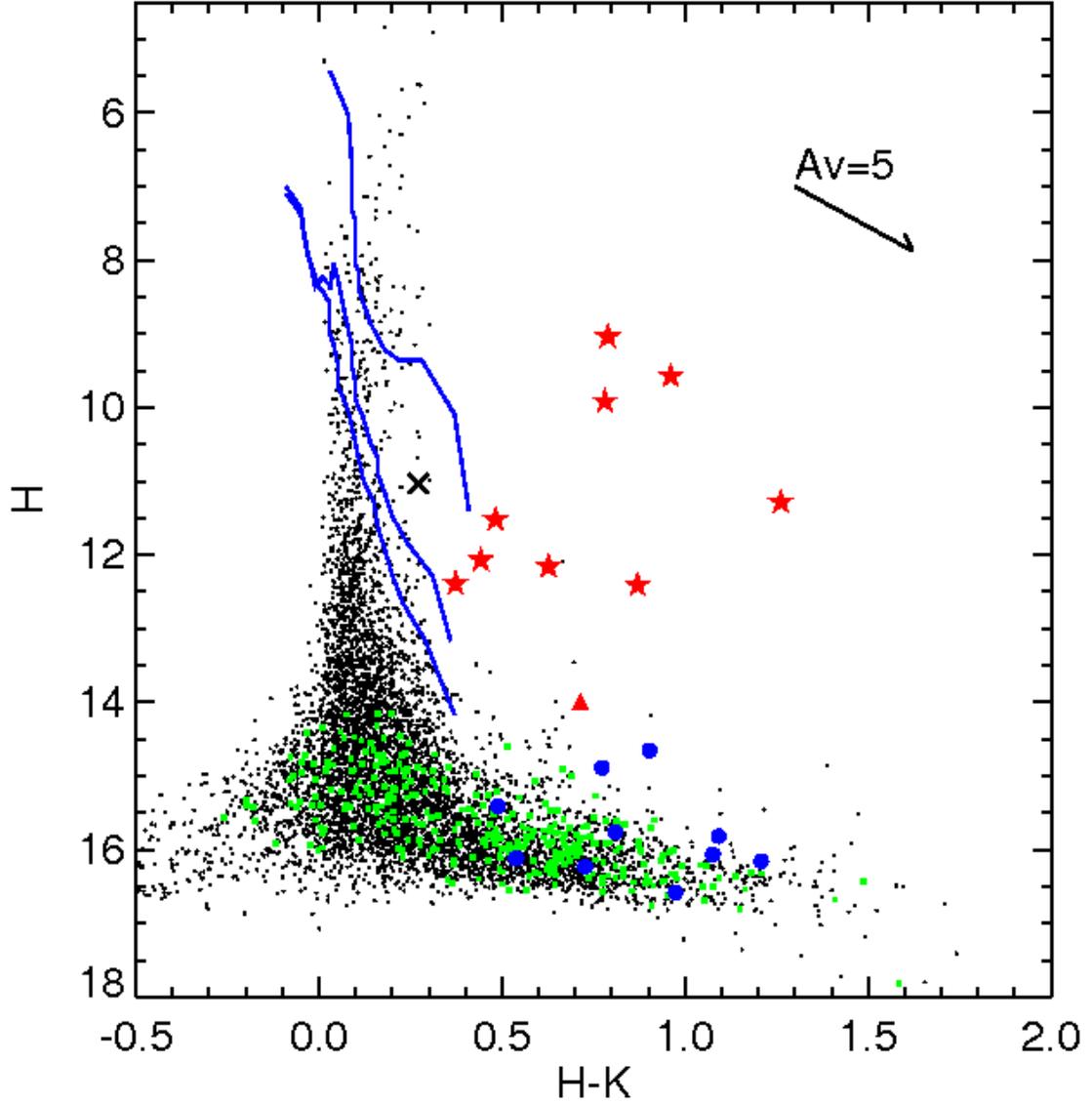}
\caption{ $H$ versus $H-K_s$ diagram for the sample, with the same
notation as earlier figures. The solid blue lines are models from
\cite{Siess-2000} at 1, 10 and 30 Myr and scaled to 210 pc. The
$\times$ symbol is the wTTS candidate (2MASS~J05060574-0646151, see
\S\ref{sec:prev}). \label{fig:hhk}}
\end{figure}

Our 10th YSO candidate is the MIPS-selected object, 050721.9-061152.
This object's SED appears in Figure \ref{fig:SED70} and shows that the
strong infrared excess starts after $\sim$8 $\mu$m (which is why this
object is not found via the IRAC-based selection method).  This kind
of SED is typical of a YSO seen edge-on.  At wavelengths shorter than
$\sim$24 $\mu$m, the thermal emission is absorbed by the disk itself
and the source is mostly seen in scattered light; this object is
indicated by the the triangles in  Figure \ref{fig:vvi}, where it is
below the  pre-main-sequence locus defined by the other members.  At
longer wavelengths, the disk becomes optically thin and the thermal
emission dominates the SED.  In Figure~\ref{fig:SED70}, we have
compared the SED of 050721.9-061152 with two well known edge-on disk
SEDs: HH~30 \citep{Burrows-1996} and HV~Tau~C \citep{Duchene-2010}. 
The flux densities have been normalized to match that of
050721.9-061152 in the $J$ band. The figure illustrates that the
global shape of the 050721.9-061152 SED, with a minimum at 8 $\mu$m,
is compatible with an edge-on disk YSO. In our color-magnitude
(Figures \ref{fig:vvi}, \ref{fig:hhk}, and
\ref{fig:i1i1m1}) and color-color (Figures \ref{fig:i1i2i3i4} and
\ref{fig:jhhk}) diagrams, this source is plotted with a red triangle.
Its location in the optical CMDs excludes its identification as simply
a debris disk or transition disk. Since photospheric optical and NIR
emissions are seen through the disk in scattered light, they are hard
to predict and broad-band observations cannot tell us much about the
nature of the central object. If it is an edge-on disk, then it joins
a relatively short list of similar objects; such objects can tell us
much about the physical properties of the occulting edge-on disk. 
Further study of this object is warranted.

Out of our entire list of ten YSO candidates, six are new discoveries.
Four of our candidates are already known as young members of IC~2118;
they were discovered by \cite{Kun-2001,Kun-2004} as classical T~Tauri
stars (cTTS) using H$\alpha$ objective prism plates and confirmed by
spectroscopy to be young. They are 050711.6-061510, 050653.5-061712,
050730.6-061060 and 050730.2-061016. For one of these,
050711.6-061510, the SED appears `disjoint' between the $I_c$- and
$J$-band observations; this is not uncommon in T~Tauri stars, and is
usually explained by the star having undergone significant variation
in between the epochs of observations.  In fact, this star is one of
three that appear in our optical observations to be nebulous and
highly variable ($>$0.2 mag) even within the timescale of our
observations.

In the spirit of the \cite{Lada-1987} and \cite{Wilking-2001} Class
I/Flat/II/III system, we fit a line to all of the  available data
between 3.6 to 24 \mum.   Eight of the IRAC-bright YSO candidates are
Class~II T~Tauri stars and one (050729.4-061637) is a flat disk. These
classifications did not change if we instead fit just 3.6 to 8 \mum. 
We did not attempt a similar fit of the edge-on-disk candidate, since
the object is seen through the disk, and the slope changes
significantly  depending on whether the MIPS points are included in
the fit.

\begin{deluxetable}
{llr@{$\pm$}rrrrr@{$\pm$}rr@{$\pm$}rr@{$\pm$}rrrrrrr@{$\pm$}r}
\rotate
\tabletypesize{\tiny}
\tablewidth{21.5cm}
\tablecaption{Measured magnitudes for Spitzer-selected YSO candidates
in IC 2118.\tablenotemark{a} \label{tab:yso}}
\tablehead{\colhead{SST Name} & \colhead{2MASS Name} & \colhead{$U$} &
\colhead{$\sigma$} & \colhead{$V$} & \colhead{$R_c$} & \colhead{$I_c$} & \colhead{$J$} & \colhead{$\sigma$} & \colhead{$H$} & \colhead{$\sigma$} & \colhead{$K_s$} & \colhead{$\sigma$} & \colhead{[3.6]} & \colhead{[4.5]} & \colhead{[5.8]} & \colhead{[8.0]} & \colhead{[24]} & \colhead{[70]} & \colhead{$\sigma$}}
\startdata
 050650.1-061908                   & 05065009-0619079 & \nodata &   \nodata & 21.59 & 19.35 & 17.43 & 13.30 &    0.027 & 12.16 &    0.026 & 11.53 &    0.025 & 10.75 & 10.38 & 10.04 &  9.52 &  6.79 &     \nodata &   \nodata\\ 
 050653.5-061712\tablenotemark{b}  & 05065349-0617123 & 20.99   &    0.102  & 17.00 & 15.20 & 13.71 & 11.16 &    0.026 &  9.91 &    0.027 &  9.13 &    0.024 &  8.20 &  7.76 &  7.31 &  6.51 &  3.27 &      0.53   &    0.044\\ 
 050711.6-061510\tablenotemark{b,c}& 05071157-0615098 & 18.91   &    0.033  & 16.29 & 15.01 & 13.85 & 13.02 &    0.026 & 11.28 &    0.027 & 10.02 &    0.029 &  8.74 &  8.11 &  7.63 &  7.07 &  4.78 &    \nodata &   \nodata\\ 
 050720.6-061612                   & 05072058-0616119 & 21.93   &    0.225  & 19.50 & 17.58 & 15.47 & 12.91 &    0.020 & 12.07 &    0.021 & 11.63 &    0.021 & 10.82 & 10.30 &  9.86 &  9.13 &  6.42 &     \nodata &   \nodata\\ 
 050721.9-061152                   & 05072188-0611524 & \nodata &   \nodata & 19.36 & 18.04 & 16.88 & 15.04 &    0.043 & 13.99 &    0.045 & 13.28 &    0.030 & 12.27 & 11.67 & 11.36 & 11.15 &  5.45 &      0.80   &    0.058\\ 
 050728.7-061326                   & 05072865-0613258 & 20.29   &    0.058  & 18.32 & 16.68 & 14.81 & 12.50 &    0.028 & 11.52 &    0.030 & 11.04 &    0.024 & 10.57 & 10.17 &  9.87 &  9.25 &  6.19 &    \nodata &   \nodata\\ 
 050729.4-061637                   & 05072937-0616368 & 21.43   &    0.152  & 19.03 & 17.55 & 16.10 & 13.71 &    0.023 & 12.41 &    0.024 & 11.54 &    0.019 & 10.27 &  9.44 &  8.80 &  7.63 &  4.33 &     \nodata &   \nodata\\ 
 050730.2-061016\tablenotemark{b,c}& 05073016-0610158 & 17.10   &    0.030  & 14.99 & 13.72 & 12.54 & 10.82 &    0.023 &  9.57 &    0.027 &  8.61 &    0.019 &  7.30 &  6.77 &  6.35 &  5.55 &  2.43 &    -0.39   &    0.030\\ 
 050730.6-061060\tablenotemark{b,c}& 05073060-0610597 & 17.58   &    0.030  & 14.74 & 13.47 & 12.21 & 10.12 &    0.025 &  9.04 &    0.029 &  8.25 &    0.015 &  7.40 &  7.02 &  6.70 &  6.13 &  3.41 &     1.20   &    0.103\\ 
 050733.7-061348                   & 05073372-0613474 & 20.74   &    0.080  & 19.19 & 17.41 & 15.39 & 13.01 &    0.020 & 12.39 &    0.026 & 12.02 &    0.019 & 11.40 & 10.92 & 10.55 &  9.71 &  7.08 &    \nodata &   \nodata\\ 
\enddata
\tablenotetext{a}{The errors in the $VRI_c$, IRAC, and MIPS-24 data are dominated by
systematic errors; they are 0.03-0.05, 0.05, and 0.04 mag, respectively (see text).}
\tablenotetext{b}{Previously-identified YSO; see
Table~\ref{tab:prevknown}.}
\tablenotetext{c}{Nebulous and highly variable ($>$0.2 mag) in the optical in our observations.}
\end{deluxetable}

Since spectral types are not known for six of the ten YSO candidates,
it is not easy to determine masses, ages, or degree of reddening.
However,  Figure~\ref{fig:jhhk} shows the $J-H$ versus $H-K_s$ diagram
for the sample, with the same notation as earlier figures.  By
comparison to the ZAMS, it can be seen that all have moderate
reddening.  The  source that does not show significant IR excess in
the $J-H$ versus $H-K_s$ diagram is located on the 1~Myr model
isochrone.  Likewise in the $H$ versus $H-K_s$ diagram,
Figure~\ref{fig:hhk}, the T~Tauri stars appear well separated from the
main sequence stars and the identified background contaminants. They
all appear redder than the 1~Myr isochrone model mostly because they
have infrared excess starting at $H$-band. If one makes the
simplifying and probably incorrect assumption that the offset is
entirely due to $A_v$, and trace the objects back to the 1 or 3 Myr
isochrone, the brightest objects are $\sim$2\Msun, and the faintest
are $\sim$0.1\Msun.

\subsection{Previously-identified YSOs not recovered by our selection}
\label{sec:prev}

We note that \cite{Kun-2004} reported other classical T~Tauri star
(cTTS) candidates in this region (see Table~\ref{tab:prevknown}), but
they are either on the southern end of the molecular cloud or
significantly off the nebular region, and are not covered by our
survey. 

\cite{Kun-2004} reported a weak line T~Tauri star (wTTS) candidate
(2MASS~J05060574-0646151) that may or may not be a true member of
IC~2118.  This object is within our surveyed region. This object does
not show any infrared excess in the IRAC bands and is not detected at
any of the MIPS bands. Therefore, our selection methods based on
infrared excesses will not recover it. However, its position in our
$R_c /R_c-I_c$ and $V / V-I_c$ CMDs is compatible with a pre-main
sequence star according to models and the position of our 9
(IRAC-selected) YSOs in the same diagram.  This object is represented
with a $\times$ symbol in optical and NIR color-color and
color-magnitude diagrams (Figures \ref{fig:vvi}, \ref{fig:jhhk},
\ref{fig:hhk}) and in Figure \ref{fig:radec}.

\subsection{Location of the YSO candidates}

\begin{figure}
\includegraphics[width=\hsize]{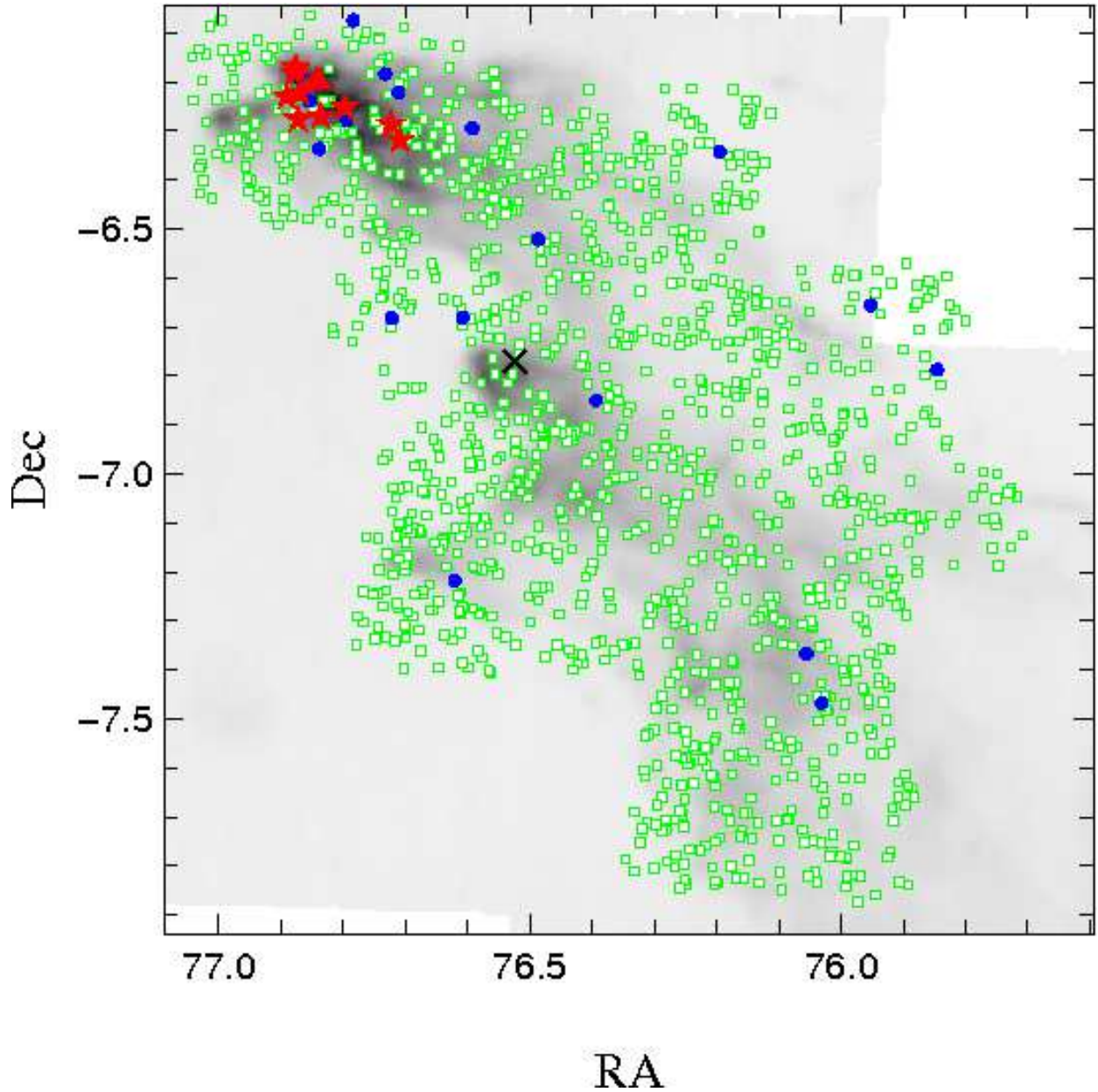}
\caption{Distribution of our YSO candidates and background galaxies
or AGN. Symbols are identical to those in Figure \ref{fig:i1i2i3i4}. 
The $\times$ symbol is the wTTS candidate (2MASS~J05060574-0646151, see
\S\ref{sec:prev}). The background gray scale image is the 160~$\mu$m
MIPS observation of IC~2118.
\label{fig:radec}}
\end{figure}

All 10 of our YSO candidates, including the edge-on-disk candidate,
are located in the head of the nebula, {\it i.e.,} G206.4-26.0, the
most massive molecular cloud \citep{Kun-2001} of IC~2118. 
Figure~\ref{fig:radec} shows the spatial distribution of the YSOs,
using the same notation as in earlier figures.  The distributions seen
here lend further support to our assertions above that the green
points are the contaminants (which are evenly distributed), the blue
points (faint IRAC-selected objects) are likely galaxies (generally
off-cloud and evenly distributed), and the IRAC- and MIPS-selected
bright objects are likely YSOs.  We expected to find YSO candidates
further South in this cloud; however, evidently the conditions do not
support substantial star formation in regions other than the
``head.''  The surface brightness of the ISM at 8, 24, 70, and 160
$\mu$m is fainter in the southern portion of the nebula compared to
the more northern portion.  The CO maps from \citet{Kun-2001} find
that the head of the nebula, G206.4-26.0, at 85 \Msun, is at least 3
times more massive than any of the other clouds.  G206.8-26.5, with 28
\Msun, houses only 2MASS~J05060574-0646151, the wTTS candidate
mentioned in the prior section, which may or may not be a true member
of the cloud and does not have an infrared excess.  At 24 \mum, the
peak surface brightness of G206.4-26.0 is $\sim$34 MJy sr$^{-1}$, and
the peak surface brightness of G206.8-26.5 is not that different at
$\sim$33 MJy sr$^{-1}$. At 160 \mum, the peak surface brightnesses are
$\sim$205 and $\sim$125 MJy sr$^{-1}$, respectively. The remaining
clouds within our map, G207.3-26.5, G207.2-27.1, and G208.1-27.5, are
2, 2, and 6 \Msun, respectively.  \citet{Kun-2001} finds them all to
be fainter, cooler, and physically smaller on the sky than either of
the other two clouds in the head of the nebula. At 24 \mum, their peak
surface brightnesses are $\sim$31, $\sim$31, and $\sim$27 MJy
sr$^{-1}$, respectively; at 160 \mum, the peaks are $\sim$77,
$\sim$30, $\sim$80 MJy sr$^{-1}$, respectively. All of this suggests
that the local conditions are different enough to significantly affect
star formation in the two locations.

\subsection{Ultraviolet excesses}

Ultraviolet (UV) excesses can also be an indication of youth (see,
e.g., \citealt{Rebull-2000} and references therein).  We have $U$-band
measurements for 9 of the known plus new candidate IC 2118 members,
including the WTTS; the two objects without $U$ measurements are
050721.9-061152 and 050650.1-061908 (see Table~\ref{tab:yso}). 
Because these objects are subject to moderate reddening, which will
significantly affect the $U$-band observations, a robust determination
of the UV excess requires a spectral type and secure determination of
$A_v$.  However, the photospheres plotted to guide the eye in
Figure~\ref{fig:seds} suggest that there are UV excesses in at least 6
of our 9 IRAC-selected YSO candidates.  Two of the largest apparent UV
excesses are in two of the stars with known types, 050730.2-061016 and
050730.6-061060 (omitting 050711.6-061510 because of the evident
stellar variation), suggesting that, when better spectral types are
obtained for the remaining objects, UV excesses may be apparent in
most of them.

\subsection{On the Distance to IC 2118}

\begin{figure}
\includegraphics[width=\hsize]{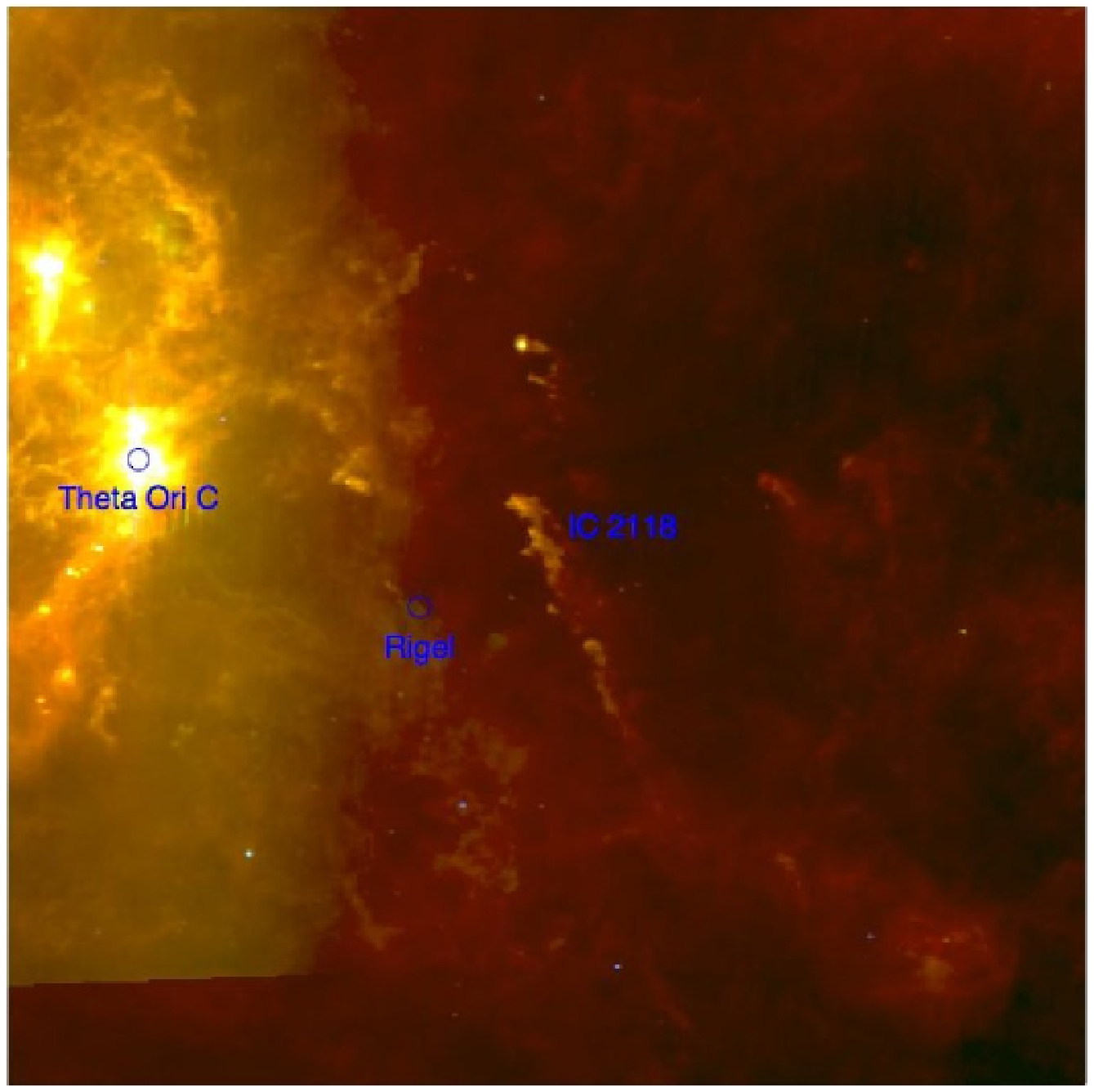}
\caption{A 3-color composite of IRAS data at 12, 25, and 100 $\mu$m,
centered on IC 2118, with a circle at the location of $\theta^1$ Ori
C and Rigel.  The region shown is $\sim$19$\arcdeg$ on a side. This figure
illustrates that IC~2118 is one of several nebulae located many
degrees from the ONC whose morphology appears ``wind-blown" by a
source roughly towards the ONC and belt stars.
\label{fig:thetaori}}
\end{figure}

\begin{figure}
\includegraphics[width=\hsize]{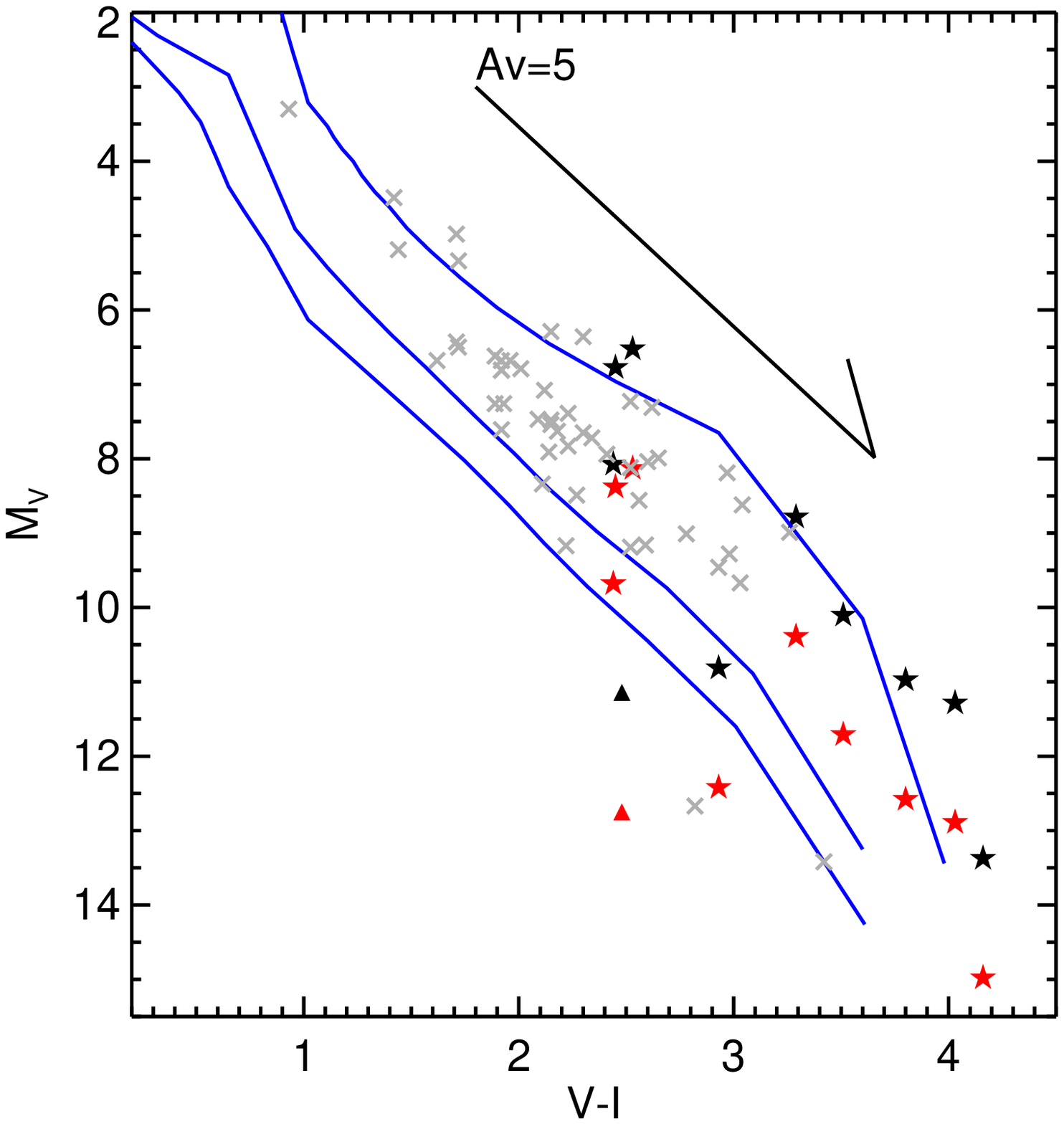}
\caption{
Optical $M_V$ {\it versus} $V-I_c$ color-magnitude diagram.  The
\cite{Siess-2000} isochrones are included (1, 10, and 30 Myr), but 
shifted to absolute $M_V$.  The black stars/triangle are our YSO
candidates, same notation as earlier figures, assuming a distance of 
440~pc, and the red stars/triangle are our YSO candidates, assuming a
distance of 210~pc.  The grey $\times$ symbols are Taurus YSOs (from
Rebull \etal\ 2010 and G\"udel et~al.\ 2007 and references therein), 
taken to be at 140~pc; \citealt{Torres-2007,Torres-2009}.  The Taurus
distribution is broad and there are many fewer IC~2118 stars, but this
distribution weakly suggests that IC 2118 is closer rather than
farther (see text). \label{fig:vvi_taurus}}
\end{figure}

At least two (perhaps) intertwined mysteries enshroud the Witch Head
-- its distance and the external source responsible for sculpting its
surface, illuminating it, and possibly triggering star-formation
within it.   There is at least a factor of two uncertainty in the
distance to IC~2118 -- 210 pc vs.\ 440 pc. Our process for identifying
YSOs towards IC 2118 does not depend strongly on the distance; the
same stars would have been selected for any distance up to the point
where the faintest objects start to blend with the background
galaxies, and all of our YSOs are considerably brighter than this
limit.   However, physical interpretation of the IC~2118 YSOs depends
on accurate determination of the distance.  For that reason, we return
to the distance question now.

Two sources (at different distances) are listed as potential
illumination and/or wind-sculpting sources for the nebula -- the
Trapezium and Rigel.  Figure~3 includes two arcs from circles
centered on the Trapezium (7.35$\arcdeg$ from IC 2118) and Rigel
(4.48$\arcdeg$ from IC 2118). We discuss these more below.

\subsubsection{Arguments in Favor of $d \sim$ 400 pc}

The most secure and accurate distance for any part of the Orion
complex is the distance derived using the VLBA for several members of
the Orion Nebula Cluster (ONC) by \cite{Menten-2007}, which is $d$ =
414 $\pm$ 7 pc.  As the most massive members of the ONC, we assume
that the Trapezium cluster stars, and in particular $\theta^1$ Ori C,
are at this distance.  Presumably, this also means that the Orion
Nebula is at this distance.  At slightly lower accuracy, the BN/KL
complex just north of the Trapezium has been determined to be at 437
$\pm$ 19 pc via VLBI measurement of water masers.  This puts the most
active current star formation and the densest molecular gas in Orion
at $\geq$ 400 pc.

As noted by \cite{Bally-1991} and expanded upon considerably by
\citet[hereafter OS98]{Ogura-1998}, there is a large population of
nebular structures surrounding the ONC, many of which have cometary
shapes or other morphology suggestive of shaping by winds or UV
photoionization.  If arrows are attached to each structure pointing
towards the apparent direction of the wind, these arrows form a
centro-symmetric pattern pointing to an origin roughly between the
Trapezium and the Orion belt stars (see Figure 2 of Bally and Figure 2
of OS98).  Both Bally and OS98 include IC~2118 as a nebula that
exemplifies this pattern.  Figure~\ref{fig:thetaori}, based on IRAS
data,  shows a wide-angle view of the gas and dust west of Orion
(centered on IC~2118), illustrating that IC~2118 is simply one of a
number of nebulae located many degrees from the ONC whose morphology
appears ``wind-blown" by a source roughly towards the ONC and belt
stars.  Detailed examination of the leading edge of the MIPS and IRAC
images of IC~2118 (Figures~\ref{fig:mosaics} and  \ref{fig:mosaic24})
supports this assertion.   The arcs in Figure~\ref{fig:mosaic24} show
that while the morphology seems to better follow the arc from Rigel,
the direction of the ``windblown'' structures in the nebula suggests
that the Trapezium is instead the exciting source, or perhaps a
structure centered near it, such as the expanding Orion-Eridanus Bubble
(see, e.g., Kun \etal\ 2001). These arguments therefore support a
distance to IC~2118 of order 400 pc.

\subsubsection{Arguments in Favor of $d \sim$ 210 pc}

Rigel ($V$ $\sim$ 0.3, spectral type B8I) is located only about
2.5$\arcdeg$ from IC~2118, compared to the Trapezium and belt stars
at more than 7$\arcdeg$ distant.  In the optical, IC~2118 is
normally described as a reflection nebula due to its blue colors and
lack of a bright rim.  The most obvious photon source -- and the one
most frequently mentioned in the literature -- for illuminating IC~2118
is Rigel.  Since, according to Hipparcos, Rigel is at d $\simeq$ 264
$\pm$ 25 pc, it follows that IC~2118 should be at a distance of order
250 pc.   

As noted by \cite{Kun-2001}, HD 32925 is located along the line of
sight to IC~2118 and appears to have significant  reddening ($A_V$ =
1.36).  This suggests it lies within or behind IC~2118.  If one adopts
the Walraven photometric distance for HD~32925 \citep{Brown-1994} of
$d\sim$ 260 pc, this places IC~2118 nearer than this.  However, the
distance to HD~32925 is not well-established.  The best spectral type
available for HD~32925 is A0III (Houk, Michigan Spectral Catalog);
given $V$ = 9.2, $M_V$ = 0.0 \citep{Schmidt-Kaler-1982} and the $A_V$, this
yields $d \sim$ 370 pc.  The original Hipparcos catalog parallax for
HD~32925 was $\pi$ = 1.69 $\pm$ 1.35 mas, hence supporting any
distance  more than about 250 pc.  The new Hipparcos reduction yields
$\pi$ = 3.66 $\pm$ 1.2 mas, or $d$ = 273 (+133, $-$66) pc.   No
firm conclusion can be drawn from these disparate estimates.

Figure~\ref{fig:vvi_taurus}  presents the $M_V$ vs.\ $V-I$ CMD,
comparing IC~2118 stars to YSOs in Taurus (with data from Rebull
\etal\ 2010, G\"udel et al.\ 2007, and references therein).  Based on
morphological grounds (e.g., the degree to which the YSOs are still
embedded in their natal gas), we expect that the IC~2118 stars might
be slightly younger than the often more physically dispersed Taurus
stars. On the other hand, based on the ratio of Class I to Class II
sources, the IC~2118 objects might be slightly older than Taurus.  In
Figure~\ref{fig:vvi_taurus}, the distribution of IC~2118 objects is
broad, but assuming a distance of 440 pc, then IC~2118 appears to be
slightly younger than Taurus. Assming a distance of 220 pc, the bulk
of the IC~2118 distribution is comparable to the bulk of the Taurus
distribution at $\sim$3 Myr. Figure~\ref{fig:vvi_taurus} thus weakly
suggests that IC~2118 is closer rather than farther.  A similar
calculation with ONC members yields similar results.


\subsubsection{No clear answer!}

\begin{figure}
\includegraphics[width=\hsize]{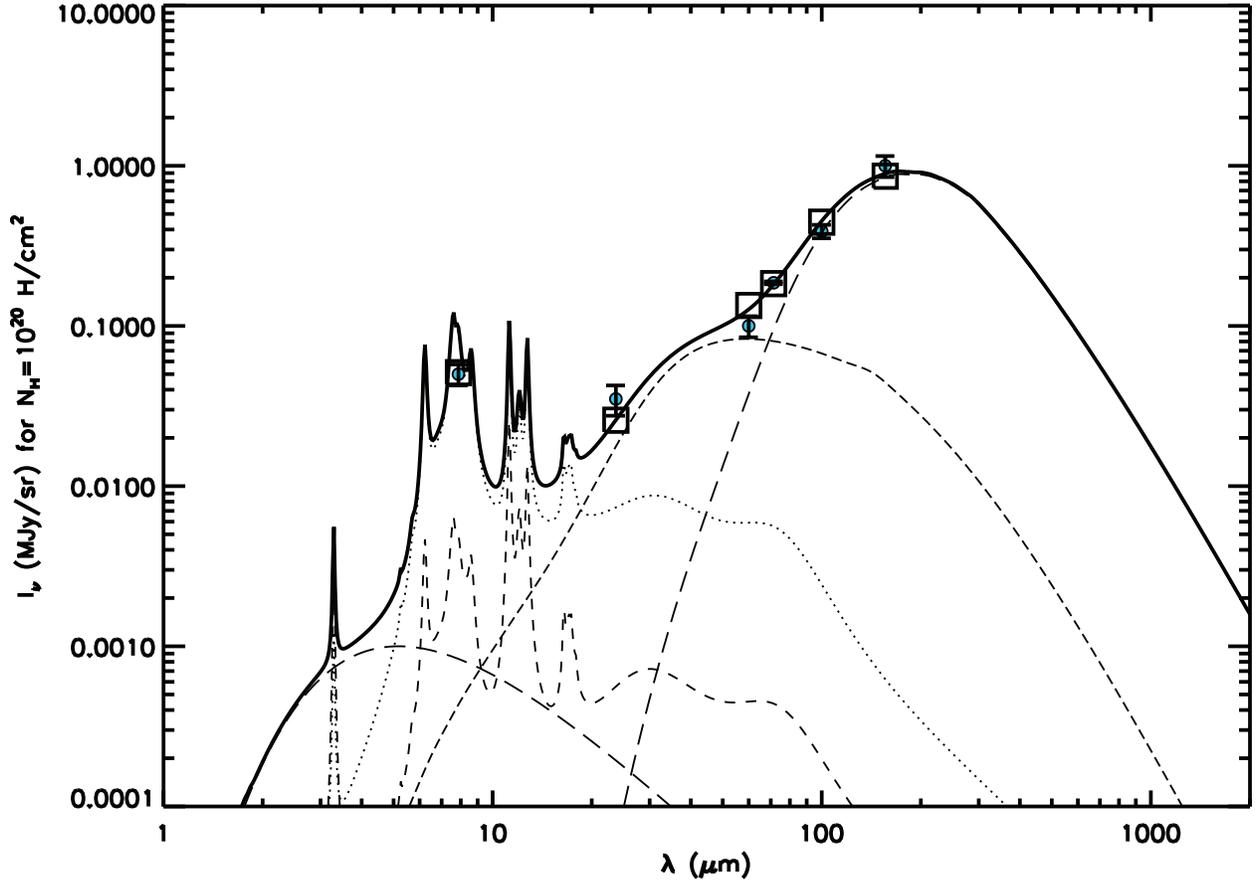}
\caption{Best model fit to the extended dust emission infrared  SED of
the IC 2118 cloud itself; see text.  The observed SED is plotted with
circles and error bars, the modeled SED with boxes. Both are
normalized to 160 \mum. The solid line is the total dust emission
spectrum; the other five lines are the different dust components (PAHs
[dotted and medium dashed lines with emission features, from the  
cations and neutrals], stochastically heated Very Small Grains
[long-dashed line, warmest blackbody curve], Big Grains at thermal
equilibrium [long-dashed cool blackbody curve]; the smooth curve that
peaks around 2-3 $\mu$m [short dashed line] is a component added in to
explain the near-IR continuum [see Flagey et al.\ 2006 and references
therein]).  The observations are well-matched by the standard
radiation field, suggesting that neither $\theta$ Ori nor Rigel are
dominant sources in illuminating IC~2118.
\label{fig:model}}
\end{figure}

It is tempting to try to use the infrared colors to set some
constraint on the distance, since after all the infrared is
reprocessing the UV starlight, and the distance between a given
source and the cloud clearly affects the radiation field bathing the
cloud. The integrated (over the entire cloud) infrared spectral
energy distribution is shown in Figure~\ref{fig:model}, where for
each band, a corresponding background outside the cloud has been
subtracted and the the images have been smoothed to the lowest
angular resolution  before the measurements. The figure includes a
fit performed using the DUSTEM model using the interstellar standard
radiation field (ISRF; Mathis et al.\ 1983) In a nutshell, the DUSTEM
model requires at least three different dust populations (PAHs, Very
Small and Big Grains) to interpret the infrared dust emissivity (for
details see Compi\'egne et al. 2008, 2010; D\'esert et al. 1990);
plus the incident radiation field that is the sum of the mean ISRF
and a stellar component diluted by the distance squared. The fact
that the observations are well-matched by the standard radiation
field immediately indicates that neither $\theta$ Ori nor Rigel are
dominant sources in illuminating the nebula, and therefore, no simple
contraint arises from the mid-IR measurements.

Unfortunately, it is our opinion that the evidence currently available
pertaining to the distance to IC~2118 is not definitive.  GAIA could
provide a definitive answer by deriving accurate parallaxes for
HD~32925 and/or for the brighter YSOs of IC~2118. A spectral type for
the diffuse optical emission from IC~2118 could establish whether the
dominant photon source is Rigel or the hotter, earlier type stars of
the ONC or belt region.

\subsection{J05080318-0617141: a high proper motion star}

While constructing 3-color images using Spitzer and POSS images, we
noticed one object projected onto the nebulosity that appeared to have
moved significantly between the epoch of the POSS image and the
current epoch; see Table~\ref{tab:propmot}.  This object,
J05080318-0617141, moves at a steady rate of $\sim$0.2 arcseconds
yr$^{-1}$ between the 1955 and 2000 epochs; at a distance of 210 pc,
this would be  $\sim$180 km s$^{-1}$. The object does not appear to
move much between the 2MASS and IRAC epochs (2000-2005); five years
may not have been enough time to see significant motion. Our optical
observations (from 2007) do not have precise enough astrometry for
this calculation.  However, this object is in the USNO-B catalog of
objects with measurable proper motion; it is object 0837-0052579,
having $\mu_{\alpha}=$134 mas yr$^{-1}$ and $\mu_{\delta}=-$86.0 mas
yr$^{-1}$, quite comparable with the net $\sim$0.2 arcseconds
yr$^{-1}$ noted above.

We were granted two nights at the Palomar 200$^{\prime\prime}$ in Jan
2007 to obtain follow-up spectroscopy of our YSO candidates.
Unfortunately, due to high velocity winds, high clouds, and poor
seeing on both of our nights, little of our data were usable.  
However, the spectrum we obtained of this relatively bright, high
proper motion object reveals it to be an M star.

We do not find an IRAC excess for this object, and it is not detected
at 24 \mum.  We have optical and near-infrared measurements: $U$=18.77,
$V$=15.98, $R_c$=14.62, $I_c$=13.19, $J$=11.68, $H$=11.08, and
$K_s$=10.85 (where the near-infrared points come from 2MASS, and the
error on all the optical points is $\sim$0.03 mags). Using the
$V/V-I_c$ diagram and the models from \cite{Siess-2000}, it is likely
to be an M2 to M5 star between 20 and 70 pc away. Assuming those
distances, its projected space motion is between $\sim$20 and 60 km
s$^{-1}$.

\begin{deluxetable}{lllllllllll}
\tablecaption{Observations of J05080318-0617141
\label{tab:propmot}}
\tabletypesize{\scriptsize}
\tablewidth{0pt}
\tablehead{
\colhead{Data set} & \colhead{position (J2000)} & \colhead{date of observation (UT)} &
\colhead{notes} } 
\startdata
POSS-1 & 05:08:02.77 -06:17:10.0 & 1955-12-15 07:50\\
POSS-2 & 05:08:03.00 -06:17:11.9 & 1985-01-20 11:26\\
2MASS & 05:08:03.19 -06:17:14.1 & 2000-10-12 09:04 & $JHK_s$ images
have a nearby ``persistence artifact"\\
IRAC & 05:08:03.18 -06:17:14.1 & 2005-03-25 00:02 & large PSF \\
our optical & 05:08:03.2 -06:17:14 & 2007-01-09 02:50& large errors on
astrometry\\
\enddata
\end{deluxetable}

\section{Discussion and Conclusions \label{sec:concl}}

By using Spitzer to search for objects with infrared excess in the
IC~2118 region, we have doubled the inventory of YSOs and YSO
candidates, in the region for which we have both IRAC and MIPS data, 
from 5 (4 CTTS and 1 WTTS) to 11 (6
new candidates selected via Spitzer colors, plus the 4 CTTS and 1 WTTS
from before).  Including the YSOs found by Kun and collaborators
outside our surveyed region, the total number of YSOs found in this
region is between 15 and 17, depending on the membership status of two
objects from \cite{Kun-2004} whose membership is questionable.  Most
of the confirmed YSOs, and all of the new objects, are found in the
``head'' of the nebula, the cloud G 206.4$-$26.0 from \cite{Kun-2001}.  
This cloud is estimated to be 85 \Msun\ total molecular mass
\citep{Kun-2001}. If we assume that all 10 YSOs located in the head of the Nebula
are legitimate
members of the cluster and on average 0.5 \Msun, then $\sim$5\% of the
cloud's mass is in stars, which is a slightly higher star formation
efficiency than could have been calculated before.    

The SEDs of all 10 of these objects are similar (Class II);
050729.4-061637 is the exception, with a flat SED that is much
different than the others. The edge-on disk candidate 050721.9-061152
has only a reasonably small excess until 24 \mum.  All 10 of these
objects are within $\sim$13$\arcmin$; assuming a distance of 210-400
pc, these objects are all within $\sim$0.8-1.6 pc of each other. 

In Perseus and other regions (\citealt{Rebull-2007} and references
therein), several small groups of likely young stars are found within
a parsec of each other.  Notably, they have a large range of SED
types, in one case (Per 6) ranging from apparent photospheres to Class
0 candidates and an apparently starless millimeter continuum core. 
The fact that the range of SED types is much smaller in the case of
IC~2118 may simply be a result of the much lower extinction towards
IC~2118, or that the stars in IC~2118 are old enough that any
differences in the timescales of very early star formation (e.g., as a
result of different initial disk masses or rotation rates) are wiped
out, or it could be telling us something about the onset of star
formation in these two different regions. There is no clear evidence
that the stars in Perseus are a result of triggered star formation,
but the literature suggests that IC~2118 was triggered. A wide range
of SED types can also be found in the cometary cloud L1251 (see, e.g.,
Lee \etal\ 2006), suggesting that perhaps triggering may not result in
a small range of SED types.

If this region is indeed triggered star formation, then we might be
able to find trends of age or mass with location. For example,
\cite{Kun-2004}  asserts that stars closer to center of the core
should be younger.  Since there are so few objects, since the distance
is uncertain, and since we do not have spectral types for most of
these objects, we can not do this rigorously; we can look at size of
infrared excess, which is often taken as a proxy for age, e.g., in the
Class 0/I/flat/II/III scheme. There is no clear correlation of various
measures of IR excess and location in the cloud. This must be, at
least in part, because so many of the SEDs are similar. 


Additional follow-up spectroscopic data are needed in the IC~2118
region to confirm or refute the YSO status of the six new objects with
infrared excesses, some of which appear to also have ultraviolet
excesses, and all of which have optical magnitudes compatible with
being YSOs in IC~2118.  The new edge-on disk candidate in particular
warrants further study, since such objects are relatively rare.

The distance to IC~2118 is still uncertain, with evidence to be found
for both $\sim$210 and $\sim$440 pc. GAIA will be needed to resolve
this mystery.

\bibliographystyle{aa}
\bibliography{mybiblio}

\begin{thebibliography}{43}
\expandafter\ifx\csname natexlab\endcsname\relax\def\natexlab#1{#1}\fi

\bibitem[{{Ballesteros-Paredes} {et~al.}(2007){Ballesteros-Paredes}, {Klessen},
  {Mac Low}, \& {Vazquez-Semadeni}}]{Ballesteros-2007}
{Ballesteros-Paredes}, J., {Klessen}, R.~S., {Mac Low}, M.-M., \&
  {Vazquez-Semadeni}, E. 2007, in Protostars and Planets V, ed. B.~{Reipurth},
  D.~{Jewitt}, \& K.~{Keil}, 63--80

\bibitem[{{Bally} {et~al.}(1991){Bally}, {Langer}, {Wilson}, {Stark}, \&
  {Pound}}]{Bally-1991}
{Bally}, J., {Langer}, W.~D., {Wilson}, R.~W., {Stark}, A.~A., \& {Pound},
  M.~W. 1991, in IAU Symposium, Vol. 147, Fragmentation of Molecular Clouds and
  Star Formation, ed. E.~{Falgarone}, F.~{Boulanger}, \& G.~{Duvert}, 11--+

\bibitem[{{Bertin} \& {Arnouts}(1996)}]{Bertin-1996}
{Bertin}, E. \& {Arnouts}, S. 1996, \aaps, 117, 393

\bibitem[{{Bessell}(1979)}]{Bessell-1979}
{Bessell}, M.~S. 1979, \pasp, 91, 589

\bibitem[{{Bessell}(1991)}]{Bessell-1991}
{Bessell}, M.~S. 1991, \aj, 101, 662

\bibitem[{{Brice{\~n}o} {et~al.}(2007){Brice{\~n}o}, {Preibisch}, {Sherry},
  {Mamajek}, {Mathieu}, {Walter}, \& {Zinnecker}}]{Briceno-2007}
{Brice{\~n}o}, C., {Preibisch}, T., {Sherry}, W.~H., {et~al.} 2007, in
  Protostars and Planets V, ed. B.~{Reipurth}, D.~{Jewitt}, \& K.~{Keil},
  345--360

\bibitem[{{Brown} {et~al.}(1994){Brown}, {de Geus}, \& {de Zeeuw}}]{Brown-1994}
{Brown}, A.~G.~A., {de Geus}, E.~J., \& {de Zeeuw}, P.~T. 1994, \aap, 289, 101

\bibitem[{{Burrows} {et~al.}(1996){Burrows}, {Stapelfeldt}, {Watson}, {Krist},
  {Ballester}, {Clarke}, {Crisp}, {Gallagher}, {Griffiths}, {Hester},
  {Hoessel}, {Holtzman}, {Mould}, {Scowen}, {Trauger}, \&
  {Westphal}}]{Burrows-1996}
{Burrows}, C.~J., {Stapelfeldt}, K.~R., {Watson}, A.~M., {et~al.} 1996, \apj,
  473, 437

\bibitem[{{Cohen}(1980)}]{Cohen-1980}
{Cohen}, M. 1980, \aj, 85, 29

\bibitem[{{Compi\'egne}(2010)}]{Compiegne-2010}
{Compi\'egne}, M. {et~al.}, 2010, A\&A, submitted

\bibitem[{{Compi\'egne}(2008)}]{Compiegne-2008}
{Compi\'egne}, M. {et~al.}, 2008, A\&A, 491, 797

\bibitem[{{D\'esert}(1990)}]{Desert-1990}
{D\'esert}, F.-X. {et~al.}, 1990, \aap, 237, 215

\bibitem[{{Duchene} {et~al.}(2009){Duchene}, {McCabe}, {Pinte}, {Stapelfeldt},
  {Menard}, {Duvert}, {Ghez}, {Maness}, {Bouy}, {Barrado y Navascues},
  {Morales-Calderon}, {Wolf}, {Padgett}, {Brooke}, \&
  {Noriega-Crespo}}]{Duchene-2010}
{Duchene}, G., {McCabe}, C., {Pinte}, C., {et~al.} 2009, \apj, in press,
  Preprint: 0911.3445

\bibitem[{{Fazio} {et~al.}(2004){Fazio}, {Hora}, {Allen}, {Ashby}, {Barmby},
  {Deutsch}, {Huang}, {Kleiner}, {Marengo}, {Megeath}, {Melnick}, {Pahre},
  {Patten}, {Polizotti}, {Smith}, {Taylor}, {Wang}, {Willner}, {Hoffmann},
  {Pipher}, {Forrest}, {McMurty}, {McCreight}, {McKelvey}, {McMurray}, {Koch},
  {Moseley}, {Arendt}, {Mentzell}, {Marx}, {Losch}, {Mayman}, {Eichhorn},
  {Krebs}, {Jhabvala}, {Gezari}, {Fixsen}, {Flores}, {Shakoorzadeh}, {Jungo},
  {Hakun}, {Workman}, {Karpati}, {Kichak}, {Whitley}, {Mann}, {Tollestrup},
  {Eisenhardt}, {Stern}, {Gorjian}, {Bhattacharya}, {Carey}, {Nelson},
  {Glaccum}, {Lacy}, {Lowrance}, {Laine}, {Reach}, {Stauffer}, {Surace},
  {Wilson}, {Wright}, {Hoffman}, {Domingo}, \& {Cohen}}]{Fazio-2004}
{Fazio}, G.~G., {Hora}, J.~L., {Allen}, L.~E., {et~al.} 2004, \apjs, 154, 10

\bibitem[{Flagey}]{Flagey-2006}
Flagey, N., Boulanger, F., Verstraete, L., Miville Deschenes, M. A.,
Noriega Crespo, A., Reach, W. T., 2006, \aap, 453, 969
  
\bibitem[{{Gautier} {et~al.}(2007){Gautier}, {Rieke}, {Stansberry}, {Bryden},
  {Stapelfeldt}, {Werner}, {Beichman}, {Chen}, {Su}, {Trilling}, {Patten}, \&
  {Roellig}}]{Gautier-2007}
{Gautier}, III, T.~N., {Rieke}, G.~H., {Stansberry}, J., {et~al.} 2007, \apj,
  667, 527

\bibitem[{{Gordon} {et~al.}(2007){Gordon}, {Engelbracht}, {Fadda},
  {Stansberry}, {Wachter}, {Frayer}, {Rieke}, {Noriega-Crespo}, {Latter},
  {Young}, {Neugebauer}, {Balog}, {Beeman}, {Dole}, {Egami}, {Haller}, {Hines},
  {Kelly}, {Marleau}, {Misselt}, {Morrison}, {P{\'e}rez-Gonz{\'a}lez}, {Rho},
  \& {Wheaton}}]{Gordon-2007}
{Gordon}, K.~D., {Engelbracht}, C.~W., {Fadda}, D., {et~al.} 2007, \pasp, 119,
  1019

\bibitem[{{G\"udel} {et~al.}(2007)}]{Guedel-2007} 
{G\"udel}, M., {et~al.} 2007, A\&A, 468, 353

\bibitem[{{Guieu} {et~al.}(2009){Guieu}, {Rebull}, {Stauffer}, {Hillenbrand},
  {Carpenter}, {Noriega-Crespo}, {Padgett}, {Cole}, {Carey}, {Stapelfeldt}, \&
  {Strom}}]{Guieu-2009}
{Guieu}, S., {Rebull}, L.~M., {Stauffer}, J.~R., {et~al.} 2009, \apj, 697, 787

\bibitem[{{Gutermuth} {et~al.}(2009){Gutermuth}, {Megeath}, {Myers}, {Allen},
  {Pipher}, \& {Fazio}}]{Gutermuth-2009}
{Gutermuth}, R.~A., {Megeath}, S.~T., {Myers}, P.~C., {et~al.} 2009, \apjs,
  184, 18

\bibitem[{{Gutermuth} {et~al.}(2008){Gutermuth}, {Myers}, {Megeath}, {Allen},
  {Pipher}, {Muzerolle}, {Porras}, {Winston}, \& {Fazio}}]{Gutermuth-2008a}
{Gutermuth}, R.~A., {Myers}, P.~C., {Megeath}, S.~T., {et~al.} 2008, \apj, 674,
  336

\bibitem[{{Jeffries} {et~al.}(2007){Jeffries}, {Oliveira}, {Naylor}, {Mayne},
  \& {Littlefair}}]{Jeffries-2007}
{Jeffries}, R.~D., {Oliveira}, J.~M., {Naylor}, T., {Mayne}, N.~J., \&
  {Littlefair}, S.~P. 2007, \mnras, 376, 580

\bibitem[{{Kun} {et~al.}(2001){Kun}, {Aoyama}, {Yoshikawa}, {Kawamura},
  {Yonekura}, {Onishi}, \& {Fukui}}]{Kun-2001}
{Kun}, M., {Aoyama}, H., {Yoshikawa}, N., {et~al.} 2001, \pasj, 53, 1063

\bibitem[{{Kun} \& {Nikolic}(2003)}]{Kun-2003}
{Kun}, M. \& {Nikolic}, S. 2003, Commmunications of the Konkoly Observatory
  Hungary, 103, 19

\bibitem[{{Kun} {et~al.}(2004){Kun}, {Prusti}, {Nikoli{\'c}}, {Johansson}, \&
  {Walton}}]{Kun-2004}
{Kun}, M., {Prusti}, T., {Nikoli{\'c}}, S., {Johansson}, L.~E.~B., \& {Walton},
  N.~A. 2004, \aap, 418, 89

\bibitem[{{Lada}(1987)}]{Lada-1987}
{Lada}, C.~J. 1987, in IAU Symp. 115: Star Forming Regions, ed. M.~{Peimbert}
  \& J.~{Jugaku}, 1--17

\bibitem[{{Lee} \& {Chen}(2007)}]{Lee-2007}
{Lee}, H.-T. \& {Chen}, W.~P. 2007, \apj, 657, 884

\bibitem[{{Lee} {et~al.}(2005){Lee}, {Chen}, {Zhang}, \& {Hu}}]{Lee-2005}
{Lee}, H.-T., {Chen}, W.~P., {Zhang}, Z.-W., {Hu}, J.-Y. 2005, \apj,
624, 808

\bibitem[{{Lee} {et~al.}(2006)}]{Lee-2006}
{Lee}, J.-E., {DiFrancesco}, J., {Lai}, S.-P., {et~al.} 2006, \apj,
648, 491

\bibitem[{{Lejeune} {et~al.}(1997){Lejeune}, {Cuisinier}, \&
  {Buser}}]{Lejeune-1997}
{Lejeune}, T., {Cuisinier}, F., \& {Buser}, R. 1997, \aaps, 125, 229

\bibitem[{{Lejeune} {et~al.}(1998){Lejeune}, {Cuisinier}, \&
  {Buser}}]{Lejeune-1998}
{Lejeune}, T., {Cuisinier}, F., \& {Buser}, R. 1998, \aaps, 130, 65

\bibitem[{{Lonsdale} {et~al.}(2003){Lonsdale}, {Smith}, {Rowan-Robinson},
  {Surace}, {Shupe}, {Xu}, {Oliver}, {Padgett}, {Fang}, {Conrow},
  {Franceschini}, {Gautier}, {Griffin}, {Hacking}, {Masci}, {Morrison},
  {O'Linger}, {Owen}, {P{\'e}rez-Fournon}, {Pierre}, {Puetter}, {Stacey},
  {Castro}, {Polletta}, {Farrah}, {Jarrett}, {Frayer}, {Siana}, {Babbedge},
  {Dye}, {Fox}, {Gonzalez-Solares}, {Salaman}, {Berta}, {Condon}, {Dole}, \&
  {Serjeant}}]{Lonsdale-2003}
{Lonsdale}, C.~J., {Smith}, H.~E., {Rowan-Robinson}, M., {et~al.} 2003, \pasp,
  115, 897

\bibitem[{{Makovoz} \& {Marleau}(2005)}]{Makovoz-2005}
{Makovoz}, D. \& {Marleau}, F.~R. 2005, \pasp, 117, 1113

\bibitem[{{Mathis}(1983)}]{Mathis-1983}
{Mathis}, J.~S. {et~al.}  1983, \aap, 128, 212

\bibitem[{{Menten} {et~al.}(2007){Menten}, {Reid}, {Forbrich}, \&
  {Brunthaler}}]{Menten-2007}
{Menten}, K.~M., {Reid}, M.~J., {Forbrich}, J., \& {Brunthaler}, A. 2007, \aap,
  474, 515

\bibitem[{{Ogura} \& {Sugitani}(1998)}]{Ogura-1998}
{Ogura}, K. \& {Sugitani}, K. 1998, Publications of the Astronomical Society of
  Australia, 15, 91

\bibitem[{{Palla} \& {Stahler}(1999)}]{Palla-1999}
{Palla}, F. \& {Stahler}, S. 1999, APJ, 525, 772

\bibitem[{{Rebull} {et~al.}(2009){Rebull}, {Guieu}, {Stauffer}, { Hillenbrand},
  {Carpenter}, {Noriega-Crespo}, { Padgett}, {Cole}, {Carey}, {Stapelfeldt}, \&
  {Strom}}]{Rebull-2009}
{Rebull}, L.~M., {Guieu}, S., {Stauffer}, J.~R., {et~al.} 2009, ApJ in prep

\bibitem[{{Rebull} {et~al.}(2000){Rebull}, {Hillenbrand}, {Strom}, {Duncan},
  {Patten}, {Pavlovsky}, {Makidon}, \& {Adams}}]{Rebull-2000}
{Rebull}, L.~M., {Hillenbrand}, L.~A., {Strom}, S.~E., {et~al.} 2000, \aj, 119,
  3026

\bibitem[{{Rebull} {et~al.}(2007){Rebull}, {Stapelfeldt}, {Evans},
  {J{\o}rgensen}, {Harvey}, {Brooke}, {Bourke}, {Padgett}, {Chapman}, {Lai},
  {Spiesman}, {Noriega-Crespo}, {Mer{\'{\i}}n}, {Huard}, {Allen}, {Blake},
  {Jarrett}, {Koerner}, {Mundy}, {Myers}, {Sargent}, {van Dishoeck}, {Wahhaj},
  \& {Young}}]{Rebull-2007}
{Rebull}, L.~M., {Stapelfeldt}, K.~R., {Evans}, II, N.~J., {et~al.} 2007,
  \apjs, 171, 447

\bibitem[{{Rieke} {et~al.}(2004){Rieke}, {Young}, {Engelbracht}, {Kelly},
  {Low}, {Haller}, {Beeman}, {Gordon}, {Stansberry}, {Misselt}, {Cadien},
  {Morrison}, {Rivlis}, {Latter}, {Noriega-Crespo}, {Padgett}, {Stapelfeldt},
  {Hines}, {Egami}, {Muzerolle}, {Alonso-Herrero}, {Blaylock}, {Dole}, {Hinz},
  {Le Floc'h}, {Papovich}, {P{\'e}rez-Gonz{\'a}lez}, {Smith}, {Su}, {Bennett},
  {Frayer}, {Henderson}, {Lu}, {Masci}, {Pesenson}, {Rebull}, {Rho}, {Keene},
  {Stolovy}, {Wachter}, {Wheaton}, {Werner}, \& {Richards}}]{Rieke-2004}
{Rieke}, G.~H., {Young}, E.~T., {Engelbracht}, C.~W., {et~al.} 2004, \apjs,
  154, 25

\bibitem[{{Schmidt-Kaler}(1982)}]{Schmidt-Kaler-1982}
{Schmidt-Kaler}, T. 1982, Bulletin d'Information du Centre de Donnees
  Stellaires, 23, 2

\bibitem[{{Siess} {et~al.}(2000){Siess}, {Dufour}, \& {Forestini}}]{Siess-2000}
{Siess}, L., {Dufour}, E., \& {Forestini}, M. 2000, \aap, 358, 593

\bibitem[{{Skrutskie} {et~al.}(2006){Skrutskie}, {Cutri}, {Stiening},
  {Weinberg}, {Schneider}, {Carpenter}, {Beichman}, {Capps}, {Chester},
  {Elias}, {Huchra}, {Liebert}, {Lonsdale}, {Monet}, {Price}, {Seitzer},
  {Jarrett}, {Kirkpatrick}, {Gizis}, {Howard}, {Evans}, {Fowler}, {Fullmer},
  {Hurt}, {Light}, {Kopan}, {Marsh}, {McCallon}, {Tam}, {Van Dyk}, \&
  {Wheelock}}]{Skrutskie-2006}
{Skrutskie}, M.~F., {Cutri}, R.~M., {Stiening}, R., {et~al.} 2006, \aj, 131,
  1163

\bibitem[{{Stansberry} {et~al.}(2007){Stansberry}, {Gordon}, {Bhattacharya},
  {Engelbracht}, {Rieke}, {Marleau}, {Fadda}, {Frayer}, {Noriega-Crespo},
  {Wachter}, {Young}, {M{\"u}ller}, {Kelly}, {Blaylock}, {Henderson},
  {Neugebauer}, {Beeman}, \& {Haller}}]{Stansberry-2007}
{Stansberry}, J.~A., {Gordon}, K.~D., {Bhattacharya}, B., {et~al.} 2007, \pasp,
  119, 1038

\bibitem[{{Stauffer} {et~al.}(2007){Stauffer}, {Hartmann}, {Fazio}, {Allen},
  {Patten}, {Lowrance}, {Hurt}, {Rebull}, {Cutri}, {Ramirez}, {Young}, {Rieke},
  {Gorlova}, {Muzerolle}, {Slesnick}, \& {Skrutskie}}]{Stauffer-2007}
{Stauffer}, J.~R., {Hartmann}, L.~W., {Fazio}, G.~G., {et~al.} 2007, \apjs,
  172, 663

\bibitem[{{Surace} {et~al.}(2004){Surace}, {Shupe}, {Fang}, {Lonsdale},
  {Gonzalez-Solares}, {Baddedge}, {Frayer}, {Evans}, {Jarrett}, {Padgett},
  {Castro}, {Masci}, {Domingue}, {Fox}, {Rowan-Robinson}, {Perez-Fournon},
  {Olivier}, {Polletta}, {Berta}, {Rodighiero}, {Vaccari}, {Stacey},
  {Hatziminaoglou}, {Farrah}, {Siana}, {Smith}, {Franceschini}, {Owen},
  {Pierre}, {Xu}, {Afonso-Luis}, {Davoodi}, {Dole}, {Pozzi}, {Salaman}, \&
  {Waddington}}]{Surace-2004}
{Surace}, J.~A., {Shupe}, D.~L., {Fang}, F., {et~al.} 2004, VizieR Online Data
  Catalog, 2255, 0

\bibitem[{{Torres} {et~al.}(2007){Torres}, {Loinard}, {Mioduszewski}, \&
  {Rodr{\'{\i}}guez}}]{Torres-2007}
{Torres}, R.~M., {Loinard}, L., {Mioduszewski}, A.~J., \& {Rodr{\'{\i}}guez},
  L.~F. 2007, \apj, 671, 1813

\bibitem[{{Torres} {et~al.}(2009){Torres}, {Loinard}, {Mioduszewski}, \&
  {Rodr{\'{\i}}guez}}]{Torres-2009}
{Torres}, R.~M., {Loinard}, L., {Mioduszewski}, A.~J., \& {Rodr{\'{\i}}guez},
  L.~F. 2009, \apj, 698, 242

\bibitem[{{van Leeuwen}(2007)}]{van-Leeuwen-2007}
{van Leeuwen}, F. 2007, \aap, 474, 653

\bibitem[{{Werner} {et~al.}(2004){Werner}, {Roellig}, {Low}, {Rieke}, {Rieke},
  {Hoffmann}, {Young}, {Houck}, {Brandl}, {Fazio}, {Hora}, {Gehrz}, {Helou},
  {Soifer}, {Stauffer}, {Keene}, {Eisenhardt}, {Gallagher}, {Gautier}, {Irace},
  {Lawrence}, {Simmons}, {Van Cleve}, {Jura}, {Wright}, \&
  {Cruikshank}}]{Werner-2004}
{Werner}, M.~W., {Roellig}, T.~L., {Low}, F.~J., {et~al.} 2004, \apjs, 154, 1

\bibitem[{{Wilking} {et~al.}(2001){Wilking}, {Bontemps}, {Schuler}, {Greene},
  \& {Andr{\'e}}}]{Wilking-2001}
{Wilking}, B.~A., {Bontemps}, S., {Schuler}, R.~E., {Greene}, T.~P., \&
  {Andr{\'e}}, P. 2001, \apj, 551, 357

\end{thebibliography}

\acknowledgements 

We wish to acknowledge all of the students who contributed their time
and energy to early reductions, analysis, discussion, and poster
papers based on this work.  They include the following, from 2005-2007:\\
From Oil City Area Senior High School (w/ T.~Spuck): D.~H.~Bowser II,
B.~R.~Ehrhart, I.~Frost, M.~T.~Heath, N.~Kelley, P.~Morton,
M.~Walentosky, S.~P.~Weiser, D.~Yeager\\
From Phillips Exeter Academy (w/ A.~Maranto): M.~T.~Greer,
J.~V.~Preis, P.~D.Weston\\
From Lincoln High School (w/ B.~Sepulveda): A.~S.~Hughes, S.~Meyer,
N.~D.~Sharma, E.~Sharma\\
From Luther Burbank High School (w/ C.~Weehler): J.~M.~Herrera

We also wish to acknowledge conversations with Gaspard Duchene 
regarding our edge-on disk candidate and fluxes for some comparison SEDs.

This work is based in part on observations made with the Spitzer
Space Telescope, which is operated by the Jet Propulsion Laboratory,
California Institute of Technology, under a contract with NASA. Support
for this work was provided by NASA through an award issued by
JPL/Caltech.  This work was also supported by the Spitzer Research 
Program for Teachers and Students.

The research described in this paper was partially carried out at the
Jet Propulsion Laboratory, California Institute of Technology, under
contract with the National Aeronautics and Space Administration.  

We wish to thank the Palomar Observatory and of course the Spitzer
staff for their assistance using the telescope. 

This research has made use of NASA's Astrophysics Data System (ADS)
Abstract Service, and of the SIMBAD database, operated at CDS,
Strasbourg, France.  This research has made use of data products from
the Two Micron All-Sky Survey (2MASS), which is a joint project of
the University of Massachusetts and the Infrared Processing and
Analysis Center, funded by the National Aeronautics and Space
Administration and the National Science Foundation.  These data are
served by the NASA/IPAC Infrared Science Archive, which is operated
by the Jet Propulsion Laboratory, California Institute of Technology,
under contract with the National Aeronautics and Space
Administration.  This research has made use of the Digitized Sky
Surveys, which were produced at the Space Telescope Science Institute
under U.S. Government grant NAG W-2166. The images of these surveys
are based on photographic data obtained using the Oschin Schmidt
Telescope on Palomar Mountain and the UK Schmidt Telescope. The
plates were processed into the present compressed digital form with
the permission of these institutions.

\end{document}